\documentclass[trackchanges,twocolumn]{aastex7}
\usepackage{multirow}
\usepackage{booktabs}
\usepackage{rotating}
\usepackage{amsmath}
\usepackage{graphics}
\usepackage{mathtools}
\usepackage{url}

\newcommand{\ts}{\thinspace}
\defcitealias{Gupta2024}{G24}

\begin{document}

\title{Resolved UV--Optical HST Imaging and Spectral Energy Distribution Modeling of Nearby BAT Active Galactic Nuclei}

\author[0000-0002-5504-8752]{Connor Auge}
\affiliation{Eureka Scientific, Inc., 2452 Delmer Street, Suite 100, Oakland, CA 94602-3017, USA}
\email[show]{connor.auge@gmail.com}  

\author[0000-0002-7998-9581]{Michael Koss}
\affiliation{Eureka Scientific, Inc., 2452 Delmer Street, Suite 100, Oakland, CA 94602-3017, USA}
\email{Mike.Koss@eurekasci.com} 

\author[0009-0007-9018-1077]{Kriti K. Gupta}
\affiliation{STAR Institute, Li\`ege Universit\'e, Quartier Agora - All\'ee du six Ao\^ut, 19c B-4000 Li\`ege, Belgium}
\affiliation{Sterrenkundig Observatorium, Universiteit Gent, Krijgslaan 281 S9, B-9000 Gent, Belgium}
\email{kkgupta@uliege.be} 

\author[0000-0001-5231-2645]{Claudio Ricci}
\affiliation{Instituto de Estudios Astrofísicos, Facultad de Ingeniería y Ciencias, Universidad Diego Portales, Avenida Ejército Libertador 441, Santiago, Chile}
\affiliation{Kavli Institute for Astronomy and Astrophysics, Peking University, Beijing 100871, People's Republic of China}
\email{claudio.ricci.astro@gmail.com} 

\author[0000-0002-3683-7297]{Benny Trakhtenbrot}
\affiliation{School of Physics and Astronomy, Tel Aviv University, Tel Aviv 69978, Israel}
\email{benny.trakht@gmail.com} 

\author[0000-0002-8686-8737]{Franz E. Bauer}
\affiliation{Instituto de Alta Investigaci{\'{o}}n, Universidad de Tarapac{\'{a}}, Casilla 7D, Arica, Chile}
\email{fbauer@academicos.uta.cl}

\author[0000-0001-7568-6412]{Ezequiel Treister}
\affiliation{Instituto de Alta Investigaci{\'{o}}n, Universidad de Tarapac{\'{a}}, Casilla 7D, Arica, Chile}
\email{etreister@academicos.uta.cl}

\author[0000-0003-2196-3298]{Alessandro Peca}
\affiliation{Eureka Scientific, Inc., 2452 Delmer Street, Suite 100, Oakland, CA 94602-3017, USA}
\affiliation{Department of Physics, Yale University, P.O. Box 208120, New Haven, CT 06520, USA}
\email{alessandro.peca@yale.edu}

\author[0000-0003-1673-970X]{Brad Cenko}
\affiliation{Astrophysics Science Division, NASA Goddard Space Flight Center, Mail Code 661, Greenbelt, MD 20771, USA}
\email{brad.cenko@nasa.gov}

\author[0000-0002-4377-903X]{Kohei Ichikawa}
\affiliation{Frontier Research Institute for Interdisciplinary Sciences, Tohoku University, Sendai 980-8578, Japan}
\email{ichikawa.waseda@gmail.com}

\author[0000-0001-7500-5752]{Arghajit Jana}
\affiliation{Department of Physics, SRM University AP, Amaravati 522240, India}
\email{arghajit.jana@mail.udp.cl}

\author[0000-0002-2603-2639]{Darshan Kakkad}
\affiliation{Centre for Astrophysics Research, University of Hertfordshire, College Lane, Hatfield, AL10 9AB, UK}
\email{d.m.kakkad@herts.ac.uk}

\author[0000-0002-7962-5446]{Richard Mushotzky}
\affiliation{Department of Astronomy, University of Maryland, College Park, MD 20742, USA}
\email{rmushotz@umd.edu}

\author[0000-0002-5037-951X]{Kyuseok Oh}
\affiliation{Korea Astronomy and Space Science Institute, Daedeokdae-ro 776, Yuseong-gu, Daejeon 34055, Republic of Korea}
\email{oh@kasi.re.kr}

\author[0000-0003-0006-8681]{Alejandra Rojas Lilay\'u}
\affiliation{Departamento de F\'isica, Universidad T\'ecnica Federico Santa Mar\'ia, Vicu\~{n}a Mackenna 3939, San Joaqu\'in, Santiago, Chile}
\email{alejandra.rojasl@usm.cl}

\author[0000-0002-1233-9998]{David Sanders}
\affiliation{Institute for Astronomy, University of Hawai`i, 2680 Woodlawn Drive, Honolulu, HI 96822, USA}
\email{sanders@ifa.hawaii.edu}

\author[0000-0003-1200-5071]{Roberto Serafinelli}
\affiliation{Instituto de Estudios Astrofísicos, Facultad de Ingeniería y Ciencias, Universidad Diego Portales, Avenida Ejército Libertador 441, Santiago, Chile}
\affiliation{INAF - Osservatorio Astronomico di Roma, Via Frascati 33, 00078, Monte Porzio Catone (Roma), Italy}
\email{roberto.serafinelli@mail.udp.cl}

\author[0000-0002-8177-6905]{Matilde Signorini}
\affiliation{European Space Agency (ESA), European Space Research and Technology Centre (ESTEC), Keplerlaan 1, 2201 AZ Noordwijk, the Netherlands}
\affiliation{INAF – Osservatorio Astrofisico di Arcetri, Largo Enrico Fermi 5, I-50125 Firenze, Italy}
\email{Matilde.Signorini@esa.int}

\author[0000-0003-3450-6483]{Alessia Tortosa}
\affiliation{INAF - Osservatorio Astronomico di Roma, Via Frascati 33, 00078, Monte Porzio Catone (Roma), Italy}
\email{alessia.tortosa@mail.udp.cl}

\author[0000-0002-0745-9792]{C. Megan Urry}
\affiliation{Physics Department and Yale Center for Astronomy \& Astrophysics, Yale University, New Haven, CT 06520, USA}
\email{meg.urry@yale.edu}


\begin{abstract}

We use high-resolution UV-to-optical imaging from the Hubble Space Telescope (HST) to construct spatially resolved spectral energy distributions (SEDs) for seven nearby ($z<0.07$) hard (14--195$\ts$keV) X-ray-selected broad-line active galactic nuclei (AGN) with $L_{\rm bol}=10^{43.26}-10^{45.34}\,\rm{erg\,s^{-1}}$. The high spatial resolution of HST, which physically resolves structures on the scale of $\sim$50$\,$pc at $z=0.05$, enables the separation of AGN and host-galaxy emission through morphological decomposition with GALFIT, yielding improved measurements of AGN properties compared to those obtained with lower-resolution Swift UV/Optical Telescope (UVOT) data. AGN UV magnitudes derived from HST imaging (e.g., F225W) can differ by more than a magnitude from those from Swift/UVOT UVM2 due to extended nuclear emission. Additionally, the inclusion of high-resolution data at longer wavelengths (e.g., F814W) can significantly affect the resulting SED fit. Comparing fits of accretion disk and extinction models using HST and Swift/UVOT data, we find significant differences in the resulting parameters, with average differences of 2.0$\,$eV in the maximum disk temperature and 2.2$\,$mag in the AGN host-galaxy extinction. These differences ultimately lead to significant changes in bolometric luminosities and X-ray bolometric corrections, with the HST-based fits yielding average increases of $\sim$0.57$\,$dex and $\sim$0.66$\,$dex respectively. This demonstrates host-galaxy contamination in unresolved UV--optical data can strongly bias SED-based estimates of disk temperatures, extinction, bolometric luminosities, and X-ray bolometric corrections in AGN. Large-area, high-resolution imaging surveys from Euclid and the Nancy Grace Roman Space Telescope will extend these techniques to much larger AGN samples, enabling uniform, high-precision SED measurements in the near-IR.

\end{abstract}


\section{Introduction} \label{sec: intro}

Active galactic nuclei (AGN) produce intense multiwavelength emission (X-ray to radio) via various physical processes. The accretion disk directly produces UV-to-optical emission, while X-rays are produced via inverse Compton scattering of UV photons in the X-ray corona. Near-infrared (NIR) to far-infrared emission is generated via the reprocessing of shorter-wavelength emission by dust at varying distances from the supermassive black hole (SMBH) \citep[for reviews, see][]{Netzer:2015:365, Hickox2018}. These multiwavelength observables are dependent on the level of obscuration surrounding the central engine, which may be driven by the SMBH accretion properties \citep{Ricci:2017:488} or by dust located relatively near the accretion disk \citep[$\sim$10 -- 100$\ts$pc;][]{Kishimoto2011A&A...536A..78K,Honig2019ApJ...884..171H,Alonso-Herrero2021A&A...652A..99A}, or at larger distances from the galactic nucleus within the host-galaxy \citep{Goulding2012,Gilli2022,Vijarnwannaluk2024MNRAS.529.3610V}. 

While detailed broadband spectral energy distributions (SEDs) can provide unique insights into these emission processes and, therefore, the intrinsic properties of the physical mechanisms driving them, it can be exceedingly challenging to accurately disentangle the AGN emission from that of the host-galaxy. host-galaxy contamination can be particularly important in low-luminosity AGN and/or heavily obscured AGN, where the UV -- optical emission from stars can dominate over that of the AGN \citep[e.g.,][]{Ho:2008:475,Simmons2011:734:121S,Peca2021}.

SED fitting software, such as CIGALE \citep{Boquien:2019:A103, Yang2022}, AGNfitter \citep{Calistro2016}, AGNFITTER-RX \citep{Martinez2024_agnfitterrx}, and SED3FIT \citep{Berta2013}, models the combined emission and disentangles the emission from the AGN and host-galaxy components through statistical means or by utilizing the AGN X-ray emission \citep{Yang:2020:740}. However, with the use of high-resolution imaging data and tools such as GALFIT \citep{Peng:2002:266}, the host-galaxy light can be effectively removed from the SED before any fitting is conducted \citep[e.g.,][]{Vasudevan:2009:1124, Shang:2011:2}. 

The BAT AGN Spectroscopic Survey \citep[BASS\footnote{https://www.bass-survey.com/};][]{Koss2017} contains 858 nearby hard X-ray selected AGN, detected using the 14--195 keV band of the Burst Alert Telescope (BAT) on board NASA's \textit{Neil Gehrels Swift} Observatory \citep{Gehrels:2004:1005}. The BASS sample was selected from the 70-month catalog \citep{Baumgartner:2013:19}; see also the 105-month catalog, \citep{Oh:2018:4} and 157-month catalogs \citep{Lien2025ApJ...989..161L}. The Swift-BAT survey selects the nearest and brightest hard X-ray selected AGN in the sky with its all-sky coverage up to Compton-thick (N$_{\rm H} = 10^{24}\ts$cm$^{2}$) obscuration levels \citep{Koss2016ApJ...825...85K}. 

Most of the sources observed by BAT in the hard X-rays have also been observed in the soft (0.5--10$\ts$keV) X-rays with the X-ray Telescope \citep[XRT;][]{Hill2005SPIE.5898..325H,Burrows2005} and the UV/Optical Telescope \citep[UVOT;][]{Roming2005SSRv..120...95R,Poole2008,Breeveld2010} on Swift. The second data release of BASS also compiled a complete optical spectroscopic analysis of all AGN in the 70-month catalog \citep{Baumgartner:2013:19}, providing an unprecedented collection of multiwavelength data with 96\% of black hole masses measured \citep{Koss:2022:1}. Detailed BASS studies have been conducted in the X-rays \citep{Ricci:2017:17,Ananna2022}, near-IR \citep{Koss:2018:214b}, mid-IR \citep{Ichikawa:2017:74,Shimizu:2017:3161,Ichikawa:2019:31}, millimeter \citep{Koss2021}, and radio \citep{Smith:2016:163,Kawamuro2023ApJS..269...24K,Magno2025}, providing measurements of both broad and narrow emission lines \citep{Oh:2022:261}, velocity dispersions \citep{Koss:2022:261}, and NIR spectral properties \citep{Gillette2026}.

Recently, \cite{Gupta2024}, hereafter \citetalias{Gupta2024}, highlighted the power of isolating the AGN emission from that of the host-galaxy by using GALFIT, fitting Swift/UVOT data with a point source and Sersic model for a sample of 236 nearby AGN from BASS. The resulting fits yielded the emission from the point-source component for each Swift/UVOT filter (V, B, U, UVW1, UVM2, and UVW2) for each source. Using these, along with Swift/XRT data that were taken simultaneously to minimize variability effects, \citetalias{Gupta2024} were able to fit the AGN SED using a variety of phenomenological models, ultimately deriving AGN characteristics such as the disk temperatures and bolometric corrections ($\kappa_{\lambda}$) for an order-of-magnitude larger sample than had previously been analyzed in this way.

In this work, we expand on the findings presented in \citetalias{Gupta2024} and highlight the power of using higher-resolution data from the {\it Hubble Space Telescope (HST)}
to more accurately remove any extended emission in the bulge of the host-galaxy that may still be unresolved in the Swift/UVOT data. We compare directly to the results in \citetalias{Gupta2024} for 7 nearby ($z < 0.07$) hard X-ray-selected AGN, selected to have the best combination of HST observations currently available for this analysis (see Section \ref{sec: data}).

This paper is organized as follows: In \S \ref{sec: data}, we describe the photometric data used in this work, along with a description of the sample. \S \ref{sec: analysis} presents the analysis of the data, describing the process of determining point source magnitudes using GALFIT for each source, along with fitting the resulting SED AGN disk models. Finally, in \S \ref{sec: results} we compare the SED fits with the results from the Swift/UVOT data in \citetalias{Gupta2024}. We assume a cosmological model with $H_{0}=70\ts\rm{km\ts s^{-1}\ts Mpc^{-1}}$, $\Omega_{\rm{M}}=0.3$, and $\Omega_{\Lambda}=0.7$ throughout this work.

\section{Data \& Sample} \label{sec: data}

\begin{figure*}
    \centering
    \includegraphics[width=0.9\textwidth]{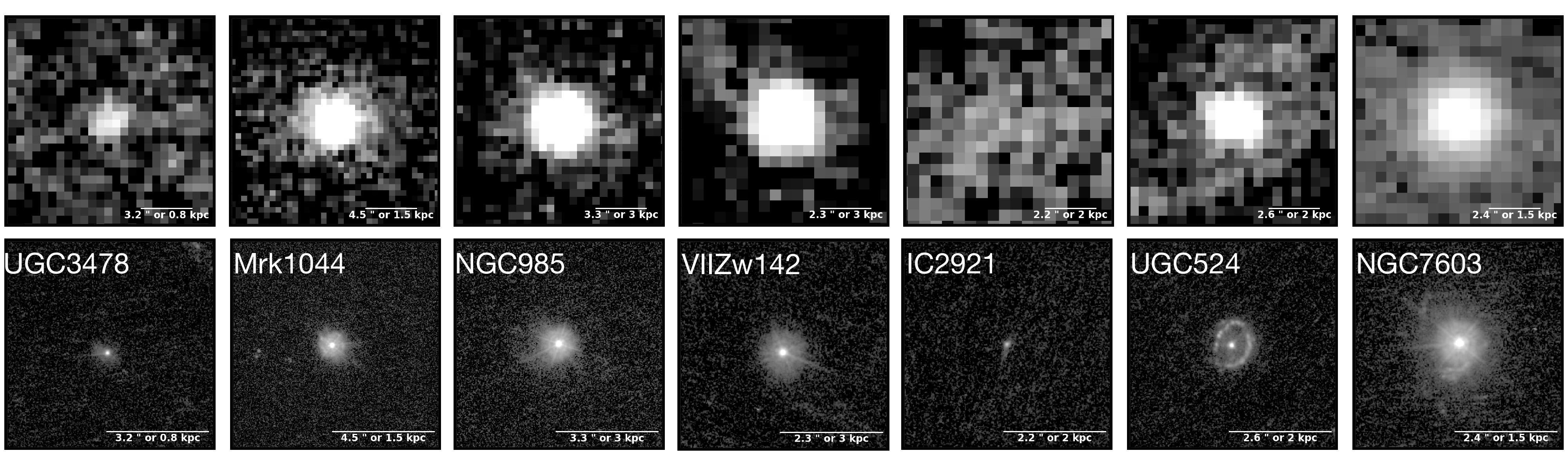}
    \caption{Cutout images of the Swift/UVOT data in the UVM2 band (top) and the HST UVIS data in the F225W band (bottom) for each source in the sample. These two filters have approximately the same central wavelength (2246.43$\ts$\AA\ and 2357.65$\ts$\AA, respectively), and the scale of each image is identical. The increased resolving power of the HST data reveals extended features that are not visible in the Swift/UVOT data. The difference in the UV magnitude of the central point source, as measured from the Swift and HST data, increases from left to right and is listed in Table \ref{tab: mags}.}
    \label{fig: cutout}
\end{figure*}

More than 506 BAT AGN have optical HST imaging available through a completed 5-year gap program in the HST ACS/WFC i-band of the entire sample at $z<0.1$ \citep{Kim:2021}, an ongoing gap program with g-band observations of 356 AGN, and WFC3/UVIS F225W data from a recent HST SNAP program. In this work, we use data from three SNAP programs: UV 225W (Koss 16241), 435W (Koss 17310), and F814W (Barth 15444). While the Swift/UVOT PSFs range from 2'' -- 3'' ($\sim$0.4 -- 4$\ts$kpc for $0.01 < z < 0.05$), comparable to typical sizes of bulge components of large galaxies, the PSFs of the HST data range between 0.075'' -- 0.08'' ($\sim$20 -- 80$\ts$pc for $0.01 < z < 0.05$). The higher resolution of the HST data enables more accurate identification of AGN emission from the rest of the host-galaxy bulge, which can be confused with lower-resolution Swift/UVOT data. The improvement in resolution between the HST and Swift/UVOT data is shown in Figure \ref{fig: cutout}. Extended features can be very clearly identified in the two right columns of Figure \ref{fig: cutout} (NGC7603 and UGC524). These features are within the central 2$\ts$kpc of the image and are not resolved by the Swift/UVOT images.

To compare the results of this analysis with those reported in \citetalias{Gupta2024}, we required the same Swift/UVOT data used in their analysis. We gather the Swift/UVOT data, along with the corresponding Swift/XRT data, from the HEASARC archive\footnote{\url{https://heasarc.gsfc.nasa.gov/cgi-bin/W3Browse/w3browse.pl}}. We selected appropriate data based on the matching observation ID listed in \citetalias{Gupta2024}, which was originally selected to determine the best possible estimates of the UV/optical fluxes at the time of \citetalias{Gupta2024} writing. 
While more recent observations are available for some sources, we chose to utilize the same data to conduct a more direct comparison with their original work. The potential effects of variability on the analysis are discussed in \S \ref{subsec: variability}.

\begin{deluxetable*}{c c c c c c c c}
\tablecaption{Sample Properties.\label{tab:sample}}
\tablehead{
\colhead{BAT ID} & 
\colhead{Counterpart Name} &
\colhead{$z$} &
\colhead{$\log M_{\rm BH}$} &
\colhead{$\log L_{2-10\,{\rm keV}}$} &
\colhead{$\log L_{\rm bol}$} &
\colhead{$\log \lambda_{\rm Edd}$} &
\colhead{Spectral Type} \\
\colhead{(1)} &
\colhead{(2)} &
\colhead{(3)} &
\colhead{(4)} &
\colhead{(5)} &
\colhead{(6)} &
\colhead{(7)} &
\colhead{(8)}
}

\startdata
1189 & NGC7603     & 0.023 & 8.59 & 43.62 & 45.34 & -1.95  & Sy1   \\
335  & UGC3478     & 0.012 & 6.06 & 42.29 & 43.26 & -0.84  & Sy1.5 \\
34   & UGC524      & 0.036 & 7.62 & 42.71 & 44.44 & -1.49  & Sy1.5 \\
134  & NGC985      & 0.043 & 8.11 & 43.76 & 44.98 & -1.32  & Sy1.5 \\
925  & VIIZw142  & 0.063 & 7.43 & 43.15 & 44.61 & -1.07  & Sy1.5 \\
130  & Mrk1044     & 0.016 & 6.45 & 42.23 & 44.04 & -0.58  & Sy1   \\
549  & IC2921      & 0.043 & 7.70 & 42.82 & 44.29 & -1.058 & Sy1.5 \\
\enddata

\tablecomments{
(1) \textit{Swift}/BAT ID \citep{Baumgartner:2013:19}; 
(2) counterpart galaxy name; 
(3) spectroscopic redshift \citep{Koss2022:261:2K}; 
(4) SMBH mass in solar masses \citep{Mejia-Restrepo2022,Koss2022:261:2K}; 
(5) logarithm of the 2--10~keV X-ray luminosity in erg\,s$^{-1}$ \citep{Ricci:2017:17}; 
(6) logarithm of the AGN bolometric luminosity from \citetalias{Gupta2024} in erg\,s$^{-1}$; 
(7) logarithm of the Eddington ratio; 
(8) optical spectral type \citep{Oh:2022:261}.
}
\label{tab:sample}
\end{deluxetable*}

We obtained X-ray data from the Swift/XRT instrument using the UK Swift Science Data Centre (UKSSDC) online data portal\footnote{\url{https://www.swift.ac.uk/swift_portal/}}. The downloaded data included cleaned and calibrated event files, automatically extracted source and background spectra, ancillary response files (ARFs), and response matrix files (RMFs), all generated using standard pipeline settings. These spectra are based on default source and background extraction regions optimized for point sources, and are suitable for spectral fitting with XSPEC \citep{xspec}. All data correspond to observations taken in Photon Counting (PC) mode.

We highlight seven BASS AGN in this work to compare how higher-resolution HST data affect the derived AGN properties and which AGN properties are most affected due to differences in resolution and variability. The sample was selected by identifying all BASS sources with the following requirements: one short (5 -- 20 s) and one long ($>$500 s) exposure in the same three HST bands, WFC3/UVIS F225W, ACS/WFC F435W, and ACS/WFC F814W (central wavelengths of $2371.15$\ts\AA, $4329.85$\ts\AA, and $8045.53$\ts\AA, respectively), a Swift/XRT observation within 4 months of the WFC3/UVIS F225W observation, and inclusion in the \citetalias{Gupta2024} analysis. The properties of the seven sources are listed in Table \ref{tab:sample} and are taken from the BASS DR2 catalog \citep{Koss:2022:1}. 
These seven sources span a wide range of black hole masses (M$_{\rm BH}$), Eddington ratios ($\lambda_{\rm Edd} = L_{\rm bol}/L_{\rm Edd}$), and redshift ($ 0.01 < z < 0.065$) as can be seen in Table \ref{tab:sample}. The HST data presented in this article were obtained from the Mikulski Archive for Space Telescopes (MAST) at the Space Telescope Science Institute. The specific observations analyzed are available at \dataset[doi: 10.17909/yxsq-td83]{https://doi.org/10.17909/yxsq-td83}.

\section{Analysis} \label{sec: analysis}

\subsection{Large Aperture Photometry} \label{subsec:HST phot}

Large-aperture photometry is first performed on each source using both Swift/UVOT and HST data in the UVM2 and F225W filters, respectively, in order to assess data quality and identify any obvious signs of variability between the Swift/UVOT and HST observations in the UV. The large aperture photometry of the Swift/UVOT data is done using the following steps outlined in the Swift/UVOT data analysis guide\footnote{https://www.Swift.ac.uk/analysis/uvot/mag.php}, using region files for the source and background defined in SAO-DS9. These region files, along with the image data, were input into the \texttt{uvotsource} command from the Swift calibration database. This provides the flux density and magnitude within an aperture with a 5'' radius defined for each image. For the HST data, we followed the procedure outlined in the WFC3 data handbook\footnote{https://HST-docs.stsci.edu/wfc3dhb/chapter-9-wfc3-data-analysis/9-1-photometry}, using the drizzled (DRZ) images from the HST data reduction pipeline. 
To make a direct comparison, we used a 5'' aperture for both Swift/UVOT and HST data, which necessarily includes some or all of the bulge of the host-galaxy along with the point-source emission of the AGN. This value is reported in the Large Aperture Photometry columns of Tabel \ref{tab: mags} for the Swift/UVOT UVM2 and HST F225W filters.

Comparing the magnitudes in this large aperture allows for a direct comparison between the HST data analyzed in this work and the Swift/UVOT data analyzed in \citetalias{Gupta2024}, without any differences in the morphological decomposition (such as PSF selection) affecting the photometric measurements. When using the same large aperture size, similar levels of contamination from the host-galaxy bulge should be present in each measurement, thus providing a clearer comparison of data quality and providing a check for the presence of variability. Differences in the measurements of the large-aperture photometry should be less sensitive to the increased resolution between the two datasets than measurements that rely on the morphological decomposition discussed in the following section.

\subsection{Morphological Decomposition} \label{subsec: GALFIT}

\begin{deluxetable*}{c c cc ccc ccc c}
\tablecaption{Large-aperture photometry and GALFIT point-source magnitudes.\label{tab:mags}}

\tablehead{
\colhead{BAT ID} &
\colhead{Counterpart} &
\multicolumn{2}{c}{Large Aperture} &
\multicolumn{7}{c}{PSF Magnitudes} \\
\colhead{} &
\colhead{} &
\colhead{UVM2} &
\colhead{F225W} &
\colhead{UVM2} &
\colhead{B} &
\colhead{V} &
\colhead{F225W} &
\colhead{F435W} &
\colhead{F814W} &
\colhead{F225W - UVM2} \\
\colhead{(1)} &
\colhead{(2)} &
\colhead{(3)} &
\colhead{(4)} &
\colhead{(5)} &
\colhead{(6)} &
\colhead{(7)} &
\colhead{(8)} &
\colhead{(9)} &
\colhead{(10)} & 
\colhead{(11)}
}

\startdata
1189 & NGC7603    & 13.90 & 15.05 & 14.44 & 14.50 & 13.96 & 15.50 & 14.27 & 13.89 & 1.06 \\
335  & UGC3478    & 18.93 & 18.69 & 19.33 & 16.94 & 16.19 & 19.76 & 16.72 & 15.59 & 0.43 \\
34   & UGC524     & 18.03 & 18.05 & 18.36 & 16.79 & 16.25 & 19.66 & 18.87 & 18.04 & 1.30 \\
134  & NGC985     & 15.17 & 14.98 & 15.22 & 15.27 & 14.87 & 15.27 & 15.04 & 15.00 & 0.05 \\
925  & VIIZw142 & 17.10 & 16.85 & 17.22 & 16.94 & 16.44 & 17.90 & 16.53 & 16.99 & 0.68 \\
130  & Mrk1044    & 15.17 & 14.98 & 15.28 & 15.27 & 15.43 & 15.97 & 15.06 & 15.09 & 0.69 \\
549  & IC2921     & 17.10 & 16.85 & 22.15 & 17.15 & 16.44 & 21.82 & 18.97 & 16.69 & -0.33 \\
\enddata

\tablecomments{
(1) \textit{Swift}/BAT ID;  
(2) host-galaxy counterpart name;  
(3) AB magnitude from 5\arcsec\ aperture photometry of UVOT UVM2;  
(4) AB magnitude from 5\arcsec\ aperture photometry of \textit{HST}/UVIS2 F225W;  
(5--7) GALFIT point-source AB magnitudes for UVM2, B, and V from \citetalias{Gupta2024};  
(8--10) GALFIT point-source AB magnitudes for \textit{HST} F225W, F435W, and F814W.
(11) The difference between the point-source magnitude in the F225W filter and UVM2 filter.
}
\label{tab: mags}
\end{deluxetable*}

As previously noted, large-aperture photometry of AGN generally leads to significant levels of contamination from the host-galaxy, even for highly luminous and unobscured sources. As shown in Figure \ref{fig: cutout}, the 5'' aperture photometry performed in the previous section includes structures that are likely associated with nuclear star formation in the host-galaxy, covering physical scales of $\sim0.2 - 1\ts$kpc at these redshifts. While the level of host-galaxy contamination is unclear when examining the Swift/UVOT images, extended emission can be clearly identified in each aperture in the higher-resolution HST imaging data. 

To obtain more accurate measurements of the AGN flux with as little host-galaxy contamination as possible, we followed the methods outlined by \citetalias{Gupta2024}. We employed GALFIT \citep{Peng:2002:266} to fit the 2D surface brightness profiles of the sources and decompose them into AGN and galaxy light. GALFIT is an image analysis tool that can model galaxy light profiles using radial distribution models such as PSF, Sersic, and Gaussian to fit the data and separate the components of the sources, including the AGN point source, disk, and bulge. 
GALFIT fits the specified functions to the 2D imaging data and provides best-fit results based on least-squares statistics. 

The primary component of the fits performed in this analysis is fitting a PSF function to the data. This model requires a PSF image for each fit. For the ACS/WFC3 F435W and F814W filters, we created PSF images for each source using the astropy photutils EPSFBuilder \citep{photutils}. This method averages bright stars in the source image to create an effective PSF. The stars were selected by visual inspection, with three to five stars chosen per image. Once cutouts are generated for each selected star, a PSF image is created using the \texttt{epsf\_builder} function.

The WFC3/UVIS F225W images rarely contained a sufficient number of bright stars to construct a suitable PSF using this method. Therefore, an empirical PSF model was utilized for this filter. The PSFs were acquired from the WFC3 instrument page at the Space Telescope Science Institute\footnote{https://www.stsci.edu/hst/instrumentation/wfc3/data-analysis/psf}. They are presented as a grid of 56 fiducial PSFs representing the spatial variation of the PSF across the detector. For each image in this work, the PSF closest to the source on the detector was used in the GALFIT routine. While this approach may be less ideal than constructing a PSF using stars in the science image, since the model PSFs do not show any potential temporal variations that may be present in the PSF, it is the recommended method by the Space Telescope Science Institute when suitable stars are not available.

Each source had a short exposure and a long exposure in each filter. The short exposure time ranged from 5 to 20$\ts$s, and the long exposure time ranged from 670 to 1040$\ts$s, depending on the filter and source. The fit was performed on both the long and short exposures for each image to assess consistency and estimate the uncertainty in the resulting PSF magnitude. Several of the long-exposure images, particularly those taken in F814W, were saturated at the center of the galaxy. For each of these images, the saturated pixels were masked and replaced with scaled pixels from the short-exposure image after properly aligning the two exposures. The results of the short-exposure fits are used for the remainder of the analysis, as they typically have better fit statistics (reduced $\chi^{2}$) and residuals, and are less likely to be saturated in the long-exposure images. However, the results of the long and short exposures are directly compared to check that the point-source magnitude derived from each fit is consistent within errors. If inconsistencies are found, the fits are re-examined and repeated, making alterations to the PSFs or fitting parameters as needed until consistent results are obtained. 

Each fit uses three components: a PSF model to determine the point-source magnitude of the AGN, a Sersic profile to model the extended bulge, and a background model.
As the errors reported in GALFIT only account for statistical uncertainties in the flux and are therefore frequently underestimated \citep{Peng:2002:266}, we adopt an iterative fitting approach to determine more accurate errors. This is discussed in more detail in Appendix \ref{sec: GALFIT errors}.
The final best-fit PSF magnitudes are presented in Table \ref{tab: mags}. All magnitudes in this table and through the paper are reported as AB magnitudes.

\subsection{AGN Variability} \label{subsec: variability}

AGN are known to vary across a wide range of wavelengths, with timescales that depend on the distance from the SMBH where the emitting material is located. Higher energy photons, such as X-ray emissions from the hot corona, vary on timescales of days to weeks, while MIR emissions from warm dust vary over months to years \citep{Sartori:2018:L34,Sartori2019ApJ...883..139S}. These timescales depend on many contributing AGN properties and can differ from one source to another. The variability in the UV is primarily driven by instabilities in the accretion disk, changes in the accretion rate, and changes in the neutral Hydrogen column density ($N_{\rm H}$) \citep[e.g.,][]{Ulrich1997ARA&A..35..445U,Dexter2011ApJ...727L..24D,Ruan2014ApJ...783..105R}. Therefore, it is necessary to examine whether any differences in the point-source magnitude measured with GALFIT using HST and Swift/UVOT data can be attributed to the source's intrinsic variability rather than to the improved resolution of the HST data. On timescales of days to weeks, AGN variability in the UV is typically on the order of 0.05 -- 0.2 magnitudes and up to 1 magnitude for timescales of months to years in extreme cases \citep{RicciTrakhtenbrot2023NatAs}. However, large magnitude changes tend to be associated with dramatic state changes or ``changing-look" AGN, which are less common than the lower-magnitude variability we might expect from the sources in this sample \citep[e.g.,][]{LaMassa:2015:144,Temple2023MNRAS.518.2938T}. NGC7603 has been identified as a changing-look AGN in \cite{Jana2025A&A...693A..35J}, though no major state changes took place over the time period of these observations, remaining a Type 1 AGN, as determined by long-term monitoring of the source. 

As there can be anywhere from 2 to 15 months between the Swift/UVOT and HST observations, it is necessary to determine whether any differences seen in the point source magnitudes between the two datasets are due to the increased resolution of the HST data, excluding extended features, or if they are simply due to intrinsic variability between the times the observations were taken. 

\begin{figure}
    \centering
    \includegraphics[width=\linewidth]{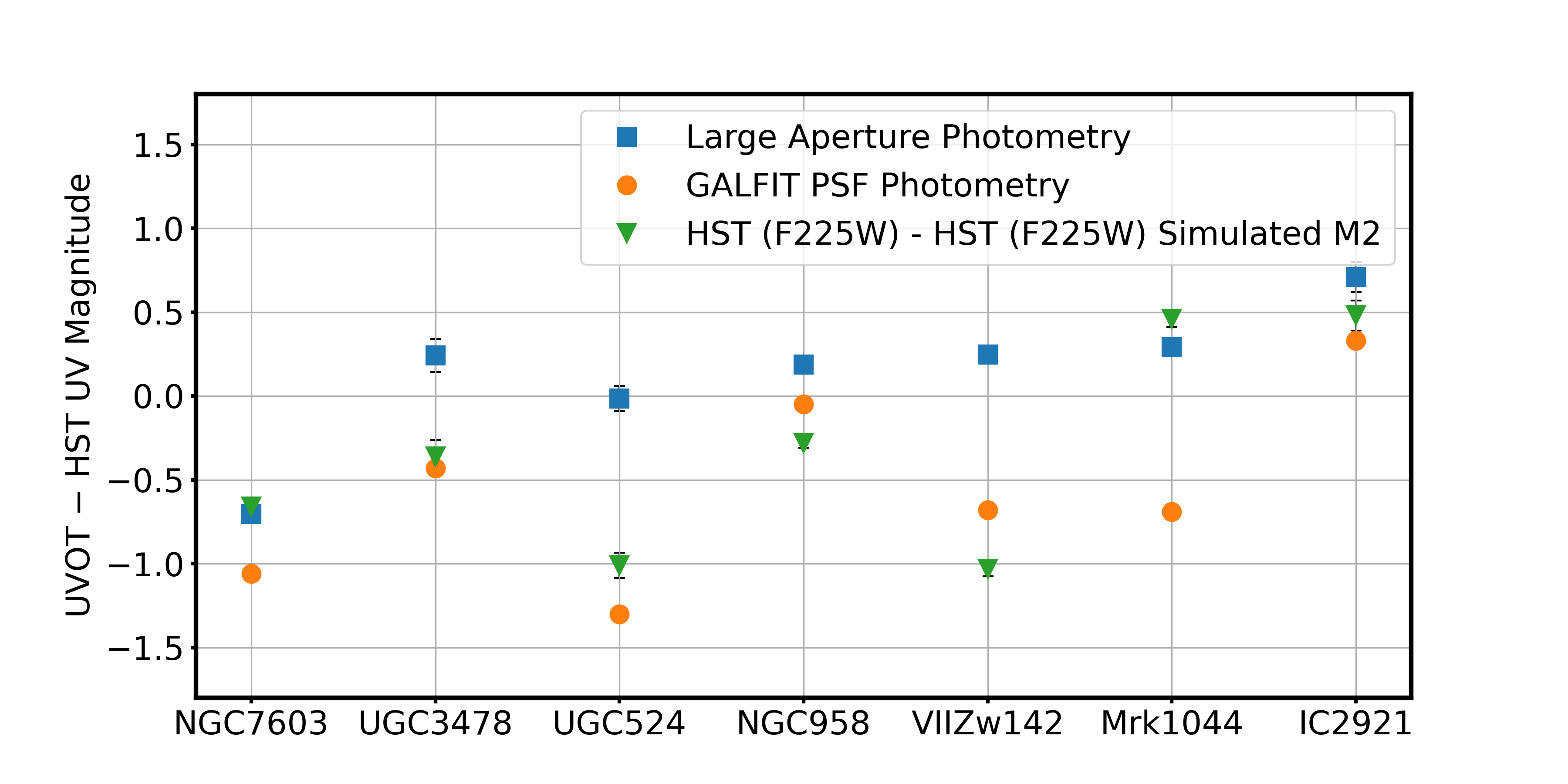}
     \caption{Magnitude differences from three different photometric measurements for each source. {\it Blue squares}: the difference in the large aperture photometry of UVM2 and F225W, {\it Orange circles}: the difference in the GALFIT point source magnitudes of UVM2 and F225W, {\it Green triangles}: the difference in the GALFIT point source magnitude of the F225W data and the GalSim-simulated UVM2 data. NGC7603 compares the F225W filter with UVOT/UVW1 instead of UVOT/UVM2 as no UVM2 data was available for this source in \cite{Gupta2024}. For all but one souce, the difference in the point-source magnitude between the HST and Swift/UVOT data is larger than the difference in the large aperture photometry.}
    \label{fig: mag comp}
\end{figure}

Figure \ref{fig: mag comp} shows a comparison between the differences in magnitudes derived from the Swift/UVOT data and the HST data obtained using both large-aperture photometry and GALFIT PSF fitting. For all but one source, the GALFIT point-source magnitude was smaller when measured from HST data than from Swift/UVOT data. This is not the case for the difference in the magnitude measured from the large-aperture photometry, which tends to be brighter when measured with HST data and generally shows smaller differences than those seen in the point-source magnitudes. The exception to this is IC2921, the faintest source in the sample in the UV, with a Swift/UVOT UVM2 magnitude of 22.15. This source was found to be brighter by $\sim$0.3 magnitudes when measured with both GALFIT point-source and large-aperture photometry using HST data. This may indicate that the differences in magnitude here are instead driven by variability rather than resolving extended features in the HST data (which do not appear to be present in Figure \ref{fig: cutout}).

The source with the second largest offset in the UV magnitude, both in the point-source magnitude and in the large-aperture photometry, is NGC 7603, which has a measured point-source magnitude in the HST F225W data that is 1.06 magnitudes fainter. However, for this source, no UVM2 data were available for the fits in \citetalias{Gupta2024}. Therefore, for comparison, we estimated the UVM2 magnitude based on the SED of the best-fit disk model found in \citetalias{Gupta2024}. While the lack of a direct UVM2 measurement could explain some of the differences seen between the derived point source magnitude of the HST F225W and the predicted UVM2 magnitude, a large discrepancy between the two would also be expected, as NGC7603 shows significant extended features in the UV imaging which are not resolved with new UVM2 imaging, and would therefore cause contamination to the lower resolution data. However, as previously mentioned, NGC7603 has shown high variability in the past, which may contribute to some of the differences seen here as well.

We checked for potential offsets in the point-source magnitudes caused by variability through an additional test using the Python program Galsim \citep{2014ascl.soft02009R}. Galsim allows high-resolution images to be used to simulate the appearance of an image at a lower resolution due to telescope/instrumental effects or to simulate the source at a higher redshift. We used HST WFC3/UVIS data to simulate how the F225W images would appear when observed with the Swift/UVOT UVM2, which has approximately the same central wavelengths (2357.65\ts\AA and 2246.43\ts\AA, respectively). This simulation was performed by convolving a simulated Swift/UVOT UVM2 PSF with the HST F225W image and adjusting the pixel scale appropriately. Determining the point source magnitude of this simulated image, following the same procedure as described in Section \ref{subsec: GALFIT}, should remove any potential variability from the equation, effectively creating a new UVOT observation that was taken at the same time as the HST data. The difference between the point source magnitude of this simulated image and that obtained using the original HST data is shown in Figure \ref{fig: mag comp}. There is good agreement between the measured point source magnitudes when comparing HST data to the observed Swift/UVOT data and to the simulated, simultaneous Swift/UVOT data for all but one source (Mrk1044), supporting the notion that this difference is likely driven by the change in data quality rather than any variability of the source for the majority of the sample.

As an additional test for the effects of variability, we constructed Swift/UVOT UVM2 light curves for all seven AGN using the available Swift monitoring data to more directly assess the role of intrinsic variability. In six of the seven sources, the peak-to-peak UV variability amplitude is smaller than the difference between the Swift/UVOT and HST F225W point-source magnitudes. NGC985 shows UV variability amplitudes larger than the point-source magnitude difference, with an average amplitude of about 0.3. However, the UVOT–HST magnitude offset is the smallest in the sample for NGC985, at 0.05 magnitudes. These results indicate that intrinsic variability cannot account for the larger magnitude differences observed in most sources and supports the interpretation that the offset arises primarily from differences in spatial resolution and host-galaxy contamination. However, for sources with small differences in the measured point source magnitudes, intrinsic source variability may play a more significant role in the measured changes between data sets. The resulting light curves and the details of their construction are discussed in more detail in Appendix \ref{sec: light curves}.

\subsection{Spectral Energy Distribution Fitting} \label{subsec: sed fits}

We now use the UV and optical point-source magnitudes from the GALFIT analysis to construct optical/UV SEDs and XSPEC-compatible PHA files for each source by converting the magnitudes to counts. 
No background PHA files were necessary for the fits, as the final magnitudes estimated by GALFIT were corrected for the background in the first step of the decomposition. Response files were generated for each filter using the defined throughput in the Python package \texttt{synphot.SpectralElement} \citep{synphot2018ascl.soft11001S}. 

The optical/UV region of the SED is fit in XSPEC v12.14.0, primarily using the same models as those used in \citetalias{Gupta2024}. These models include the DISKPN model, which is a thermal accretion disk model comprising multiple blackbody components \citep[e.g.,][]{Mitsuda1984PASJ...36..741M,Makishima1986ApJ...308..635M}. This model has been shown to successfully describe the optical/UV disk emission in many AGN sources \citep[e.g.,][]{Malkan1982ApJ...254...22M,Malkan1983ApJ...268..582M}. The ZDUST model accounts for extinction due to dust grains \citep{Pei1992ApJ...395..130P} in the optical and UV bands associated with the Milky Way (ZDUST$_{\rm MW}$) and the host-galaxy of the AGN (ZDUST$_{\rm HG}$). The convolution model ZASHIFT shifts the disk spectrum to the redshift of the respective source. Finally, we utilize the ZTBABS model to include photoionization absorption of the AGN disk at higher energies ($\sim$0.1$\ts$keV). The final combination of models is:

\begin{multline}
        \rm{ZDUST_{MW}} \times \rm{ZDUST_{HG}} \\
        \times (\rm(ZTBABS) \times \rm{ZASHIFT} \times {DISKPN})
\end{multline}

For the two reddening components, we adopted a Milky Way (MW) extinction curve (method = 1) and R$_{V}$ = 3.08. For dust extinction due to our galaxy, we fixed the $E(B-V)$ values estimated by \cite{Schlegel1998ApJ} for each source. The redshifts for the two components were fixed at 0.0 for the MW and at the spectroscopically measured redshift of the source for the host-galaxy component. The inclusion of the ZTBABS model differs from the fits conducted by \citetalias{Gupta2024}, who did not include any photoionization absorption. To see how the inclusion of this model might affect the results, we fit each source with and without this component and find that the changes in the resulting AGN properties, such as the intrinsic bolometric luminosity, are within the expected errors. However, including this model makes the observed SED of the AGN disk more realistic, removing any significant peaks at high energies where they would not be expected to be observable. The change in the observed SED due to this model's inclusion is exclusively in the EUV, where no currently data exist in either HST or Swift/UVOT; therefore, the extent of the disk emission in this region cannot be directly measured.

For the DISKPN component, the free parameters are the maximum temperature of the disk ($kT_{\rm max}$; in keV) and the host-galaxy extinction $E(B-V)$. Ideally, the normalization parameter would also be set as a free parameter; however, due to the limitation of 3 HST photometry points per source (compared to the 4 -- 5 data points used in \citetalias{Gupta2024}), only two free parameters are possible. In contrast, \citetalias{Gupta2024} was able to keep the normalization free, as well as the maximum disk temperature and host-galaxy dust extinction. We fix the normalization in the DISKPN model, which is defined as 

\begin{equation}
    K_{\rm uvo} = M_{\rm BH}^{2}cos(i)/(D^{2} \times \beta^{4})
\end{equation}

\noindent where $M_{\rm BH}$ is the mass of the central SMBH in solar masses, $i$ is the inclination angle of the disk, $D$ is the distance to the source in kpc, and $\beta$ is the color to effective temperature ratio. While using this definition to determine the normalization for a fit and fixing it, rather than leaving it as a free parameter, has been done in the past \citep[e.g.,][]{Vasudevan:2009:1124}, \citetalias{Gupta2024} presented a detailed discussion on why this approach should be done cautiously. When using this equation, the inclination is usually assumed to be zero, and the color-to-effective temperature ratio is set to unity. \citetalias{Gupta2024} showed that fixing the normalization parameter can drastically affect the overall SED fitting, as the only parameter left to vary is the disk temperature. Specifically, fixing the normalization in this way causes the resulting fits to produce higher values of the X-ray bolometric correction ($\kappa_{2-10}$) by a factor of $\sim$1.3. \citetalias{Gupta2024} argued that using a fixed value for normalization, assuming a common inclination angle and color-to-effective temperature ratio for a large sample, is likely to severely bias the final results.

\begin{figure*}
    \centering
    \includegraphics[width=0.85\linewidth]{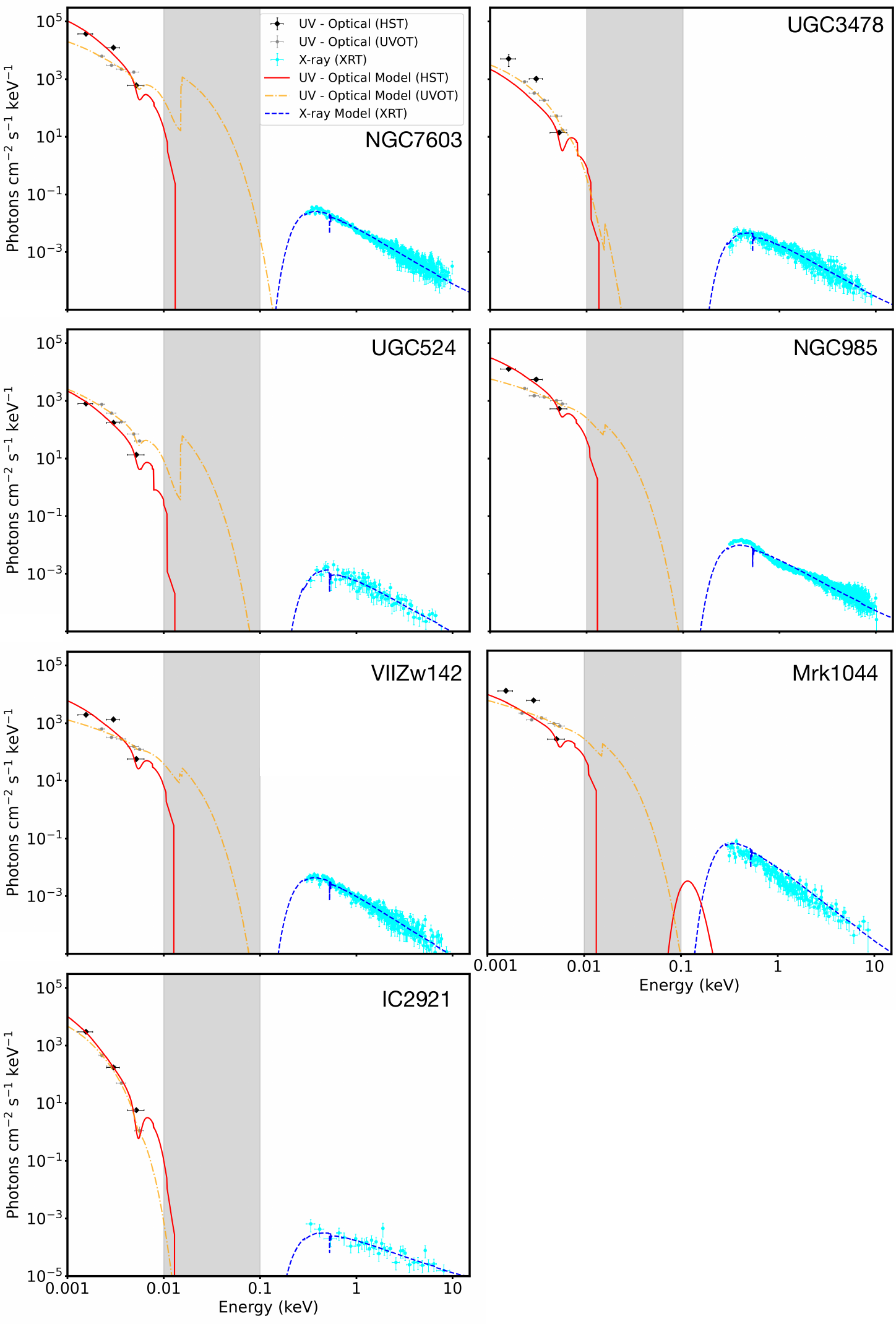}
    \caption{The optical/UV SEDs and the XRT data fit with the XSPEC models as described in section \ref{subsec: sed fits}. The black data points and red line show the HST data and fit, while the gray data points and orange line show the Swift/UVOT data and fits from \citetalias{Gupta2024}. The models fit to the HST data include a photoionization absorption component that was not included in \citetalias{Gupta2024}. This component affects the emission in the EUV that is unconstrained in both data sets (shown in in gray).}
    \label{fig: SEDs1}
\end{figure*}


However, given the lack of a fourth data point that would allow us to retain the normalization as a free parameter, we use equation 1, adopting more realistic values for the color-to-effective-temperature ratio rather than assuming $\beta \approx 1$, as was done in \cite{Vasudevan:2009:1124}. The study of narrow-line Seyfert 1 (NLS1) galaxies, which tend to have lower accretion rates, can have a color temperature correction as large as $\sim2.4$, while \cite{Davis2019} found an average correction factor of $\sim1.6$ by fitting synthetic spectral models for thin accretion disks. 

We explore fits with fixed normalization values determined with several different values for the correction factor ranging from 1, as was originally done in \cite{Vasudevan:2009:1124}, to 2.4, in order to explore how this value, and therefore the normalization, affects the both the quality of the fit and the resulting best-fit parameters. Larger the correction factors result in the normalization being fixed to a lower value, as can be seen in equation 2. Therefore, lower values of $\beta$ tended to improve the quality of the fits, and provide more reasonable disk temperatures. This is particularly true for UGC3478 and Mrk1044, the two sources with the black hole masses that are more than an order of magnitude lower than the rest of the sample. This low black hole mass drove down the normalization, so a lower correction factor was needed for these two sources, being best fit with $\beta$ values consistent with \cite{Vasudevan:2009:1124}.

The fit for the X-ray data is also performed using the method of \citetalias{Gupta2024}, with the same models and free parameters. The same binning process for the X-ray data was used, along with the cstat method to assess fit quality. The model used to fit the Swift/XRT data is defined as: (TBABS$_{\rm Gal}$ × ZPHABS × CABS × PEXRAV). This model should account for the primary X-ray emission, reflection by optically thick neutral circumnuclear material, galactic absorption, and absorption by neutral material via photoelectric absorption and Compton scattering.

The fits for the UV--optical and the X-rays are done independently of one another, matching the fitting process of \citetalias{Gupta2024}. Because some of the initial fits yielded disk temperatures larger than physically expected, the maximum disk temperature was limited to 0.1-10$\ts$eV, and the fit began with a Gaussian prior centered on the temperature originally found in \citetalias{Gupta2024}. The host-galaxy extinction was also initiated with a Gaussian prior centered on the value found from \citetalias{Gupta2024}. The best fits to both the X-ray and UV data (HST and UVOT) are shown in Figure \ref{fig: SEDs1}. 
It is important to note that some of the differences in each fit may also be attributed to variability in the obscuring column density in the X-rays ($N_{\rm H}$), as the XRT spectrum used in our fits is not the same as that used in \citetalias{Gupta2024}, but is the spectrum taken closest in time to the HST UV data.

\section{Results} \label{sec: results}
In the previous sections, we explained in detail the steps we followed to systematically construct and fit the optical-to-X-ray SEDs of a sample of seven unobscured AGN using new, high-resolution HST data, enabling us to disentangle AGN and host-galaxy emission more fully than previously possible. In this section, we focus on the outputs of our SED fitting and compare them with those reported by \citetalias{Gupta2024} using Swift/UVOT data. Figure \ref{fig: SEDs1} show the results of our fits using HST, along with those from \citetalias{Gupta2024}, and Table \ref{tab:fitting results} lists the best-fit parameters and the integrated luminosities for each fit. The bolometric luminosity is defined as $L_{\rm X} + L_{\rm uvo}$, while $L_{\rm uvo}$ and $L_{\rm X}$ are the integrated luminosities under the best fit model between $10^{-7} - 0.1\ts\rm{keV}$ and $0.1 - 500\ts\rm{keV}$, respectively. The bolometric luminosity of an AGN is a key diagnostic tool for understanding the system's physical processes, enabling determination of the Eddington ratio for each source when the SMBH mass is known.

\begin{deluxetable*}{ccccccccc}
\tablecaption{Results of XSPEC fitting comparing HST and Swift/UVOT \label{tab:fitting_results}}
\tablehead{
\colhead{BAT ID} &
\colhead{Counterpart Name} &
\colhead{kT$_{\rm max}$} &
\colhead{E(B--V)} &
\colhead{log $L_{\rm UV}$} &
\colhead{log $L_{\rm 2-10keV}$} &
\colhead{log $L_{\rm Bol}$} &
\colhead{$\kappa_{2-10\ts keV}$} &
\colhead{$\alpha_{\rm ox}$} \\
\colhead{} &
\colhead{HST / $\Delta$\citetalias{Gupta2024}} &
\colhead{HST / $\Delta$\citetalias{Gupta2024}} &
\colhead{HST / $\Delta$\citetalias{Gupta2024}} &
\colhead{HST / $\Delta$\citetalias{Gupta2024}} &
\colhead{HST / $\Delta$\citetalias{Gupta2024}} &
\colhead{HST / $\Delta$\citetalias{Gupta2024}} &
\colhead{HST / $\Delta$\citetalias{Gupta2024}} \\ 
\colhead{(1)} &
\colhead{(2)} &
\colhead{(3)} &
\colhead{(4)} &
\colhead{(5)} &
\colhead{(6)} &
\colhead{(7)} &
\colhead{(8)} &
\colhead{(9)}
}
\startdata
1189 & NGC7603 & 1.31 / $-4.6$ & 0.28 / 0.03    & 45.47 / 0.18    & 43.56 / $-0.07$ & 45.49 / 0.15  & 85.87 / 33.89  & $-1.50$ / $0.02$ \\
335 & UGC3478  & 9.98 / 8.9    & 0.59 / 0.48    & 43.95 / 1.08    & 42.30 / 0.01    & 43.99 / 0.73  & 49.29 / 40.92  & $-1.22$ / $-0.10$ \\
34 & UGC524   & 1.92 / $-1.8$ & 0.26 / $-0.04$ & 44.20 / $-0.20$ & 42.80 / 0.09    & 44.26 / -0.18 & 28.36 / -25.21 & $-1.30$ / $0.27$ \\
134 & NGC985  & 2.43 / $-1.8$ & 0.20 / 0.18    & 45.40 / 0.63    & 43.67 / $-0.08$ & 45.43 / 0.45  & 57.67 / 41.12  & $-1.45$ / $-0.13$ \\
9252 & VIIZw142  & 5.81 / 1.9    & 0.34 / 0.29    & 45.56 / 1.05    & 43.33 / 0.18    & 45.57 / 0.96  & 172.7 / 145.1  & $-1.50$ / $-0.08$ \\
130 & Mrk1044  & 10.0 / 5.5    & 0.20 / 0.17    & 44.74 / 0.86    & 42.82 / 0.39    & 44.79 / 0.75  & 92.62 / 57.28  & $-1.23$ / $0.25$ \\
549 & IC2921  & 3.87 / 3.3    & 0.86 / 0.67    & 45.39 / 1.35    & 42.87 / 0.05    & 45.40 / 1.11  & 336.2 / 307.7  & $-1.81$ / $-0.67$ \\
\enddata
\tablecomments{
(1) Swift-BAT ID.  
(2) Counterpart Name
(3) Maximum disk temperature of the best-fit model (eV) using HST.  
(4) Dust extinction from the best-fit model (mag).  
(5) Logarithm of the UV luminosity (erg\,s$^{-1}$).  
(6) Logarithm of the X-ray luminosity (erg\,s$^{-1}$).  
(7) Logarithm of the bolometric luminosity ($L_{\rm bol} = L_{\rm X-ray} + L_{\rm uvo}$).  
(8) X-ray bolometric correction $\kappa_{2-10\rm keV} = L_{\rm bol}/L_{2-10\rm keV}$.  
(9) $\alpha_{\rm OX}$ determined from $L_{\rm 2keV}$ and $L_{2500}$.
The second value in each column is the difference between the best-fit value found in this work and that found in \citetalias{Gupta2024} using the Swift/UVOT data.
}
\label{tab:fitting results}
\end{deluxetable*}

\begin{figure}
    \centering
     \includegraphics[width=\linewidth]{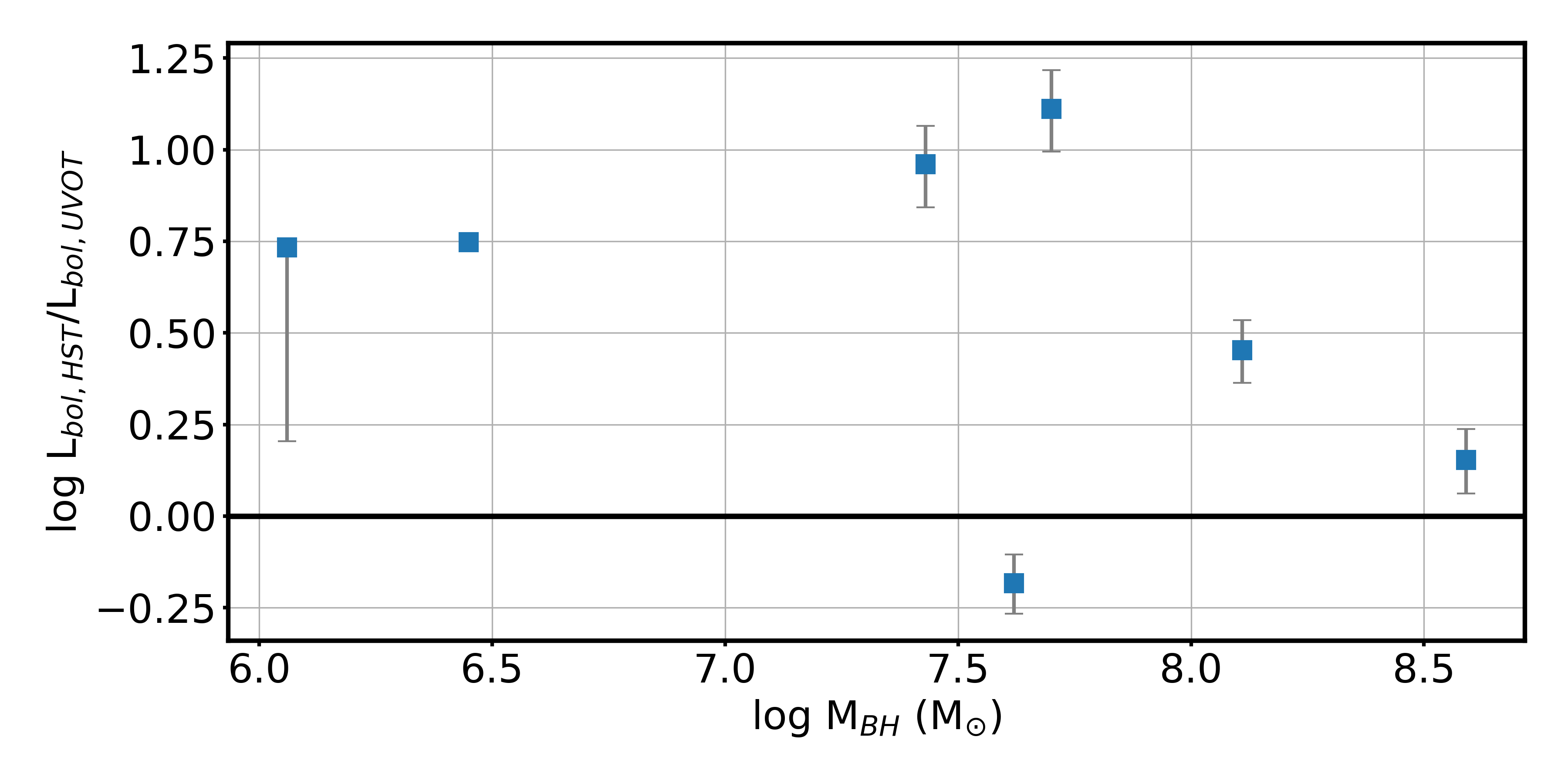}
    \caption{The ratio of the bolometric luminosity determined from the best fit SEDs in this work using the HST data in the UV--optical, and the bolometric luminosity from \citetalias{Gupta2024} using Swift/UVOT data.}
    \label{fig: Lbol_comp}
\end{figure}

\begin{figure}
    \centering
    \includegraphics[width=\linewidth]{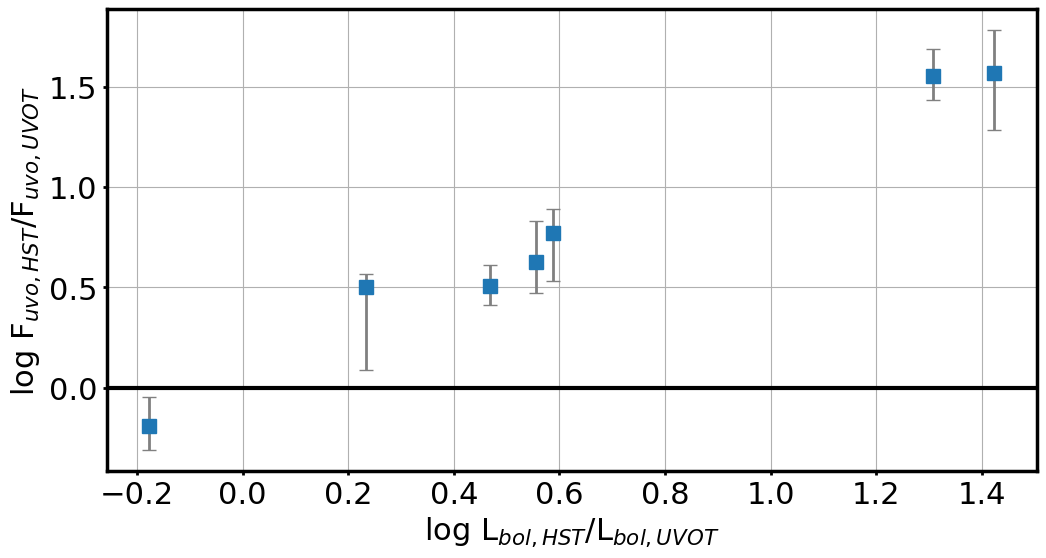}
    \includegraphics[width=\linewidth]{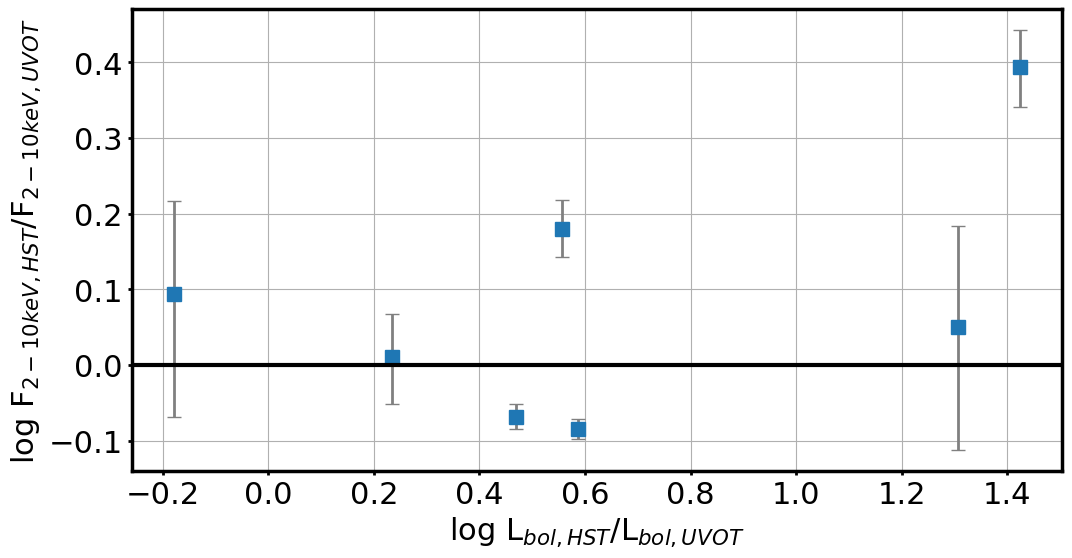}
    \caption{Top: The ratio of the flux of the UV/Optical disk model ($1\times10^{-7} - 0.1\rm{keV}$) determined from the best fit SEDs in this work using the HST data to the best UV/Optical flux found in \citetalias{Gupta2024} as a function of the difference in bolometric luminosites. Bottom: Same as the top but comparing the 2 -- 10$\ts$keV flux found in this work to that found in \citetalias{Gupta2024}.}
    \label{fig:Fuvo_Fx}
\end{figure}

Figure \ref{fig: Lbol_comp} shows how the bolometric luminosity, derived from SED fitting with the HST data, differs from the luminosity found by fitting the Swift/UVOT data in \citetalias{Gupta2024} for each source and as a function of SMBH mass. The difference in bolometric luminosity can be as high as 1 dex, with 6 of the sources showing higher values when fit to the HST data, and only one showing lower bolometric luminosity. Figure \ref{fig:Fuvo_Fx}, shows how both the total UV/optical flux and the X-ray flux change between the two fits as a function of the change in bolometric luminosity. A clear correlation is observed between changes in the UV/optical flux and the bolometric luminosity. In contrast, no clear trend is observed when comparing these changes to variations in the X-ray flux. This highlights that the change in bolometric luminosity is driven by changes in the fits to the HST data in the UV/optical, rather than by any change in the X-ray contribution, which may be driven by variability.

To test the correlation seen in the top panel of Figure \ref{fig:Fuvo_Fx}, we compute the Spearman rank correlation coefficient between $\Delta$(UV--optical flux) and $\Delta L_{\rm bol}$, obtaining $\rho$ = 1.0 for the nominal values (exact two-sided p = $4\times10^{-4}$ for n = 7). To account for asymmetric measurement uncertainties, we perform 10,000 Monte Carlo realizations, drawing each source from its quoted error distribution. The resulting distribution of correlation coefficients has a median $\rho$ = $0.82^{+0.04}_{-0.11}$, with 100$\%$ of realizations yielding $\rho >$ 0. The corresponding two-sided p-value for $\rho\approx$ 0.82 and n = 7 is p $\approx$ 0.02, indicating that the observed trend is unlikely to arise from random rank ordering.

As shown in Figure \ref{fig: SEDs1}, the exact difference between the UV and optical SED is dependent on the combination of fitting parameters. The disk temperature and host-galaxy dust extinction, as well as the model normalization—which had to be left as a fixed parameter in the fits to the HST data—affect this change in different ways while also being degenerate with one another, particularly without extensive wavelength coverage in the SED data.

\subsection{Disk Temperature}
One of the free parameters for each fit and one of the most thoroughly analyzed physical parameters in \citetalias{Gupta2024} is the maximum disk temperature. The standard accretion disk model suggested by \cite{Shakura:1973:337} considers a geometrically thin and optically thick structure for the disk with a temperature gradient, such that higher temperatures are expected in closer proximity to the SMBH, extending up to the innermost stable circular orbit. Our SED-fitting analysis uses a multi-temperature accretion disk model, allowing us to determine the maximum disk temperatures. 

\begin{figure}
    \centering
    \includegraphics[width=\linewidth]{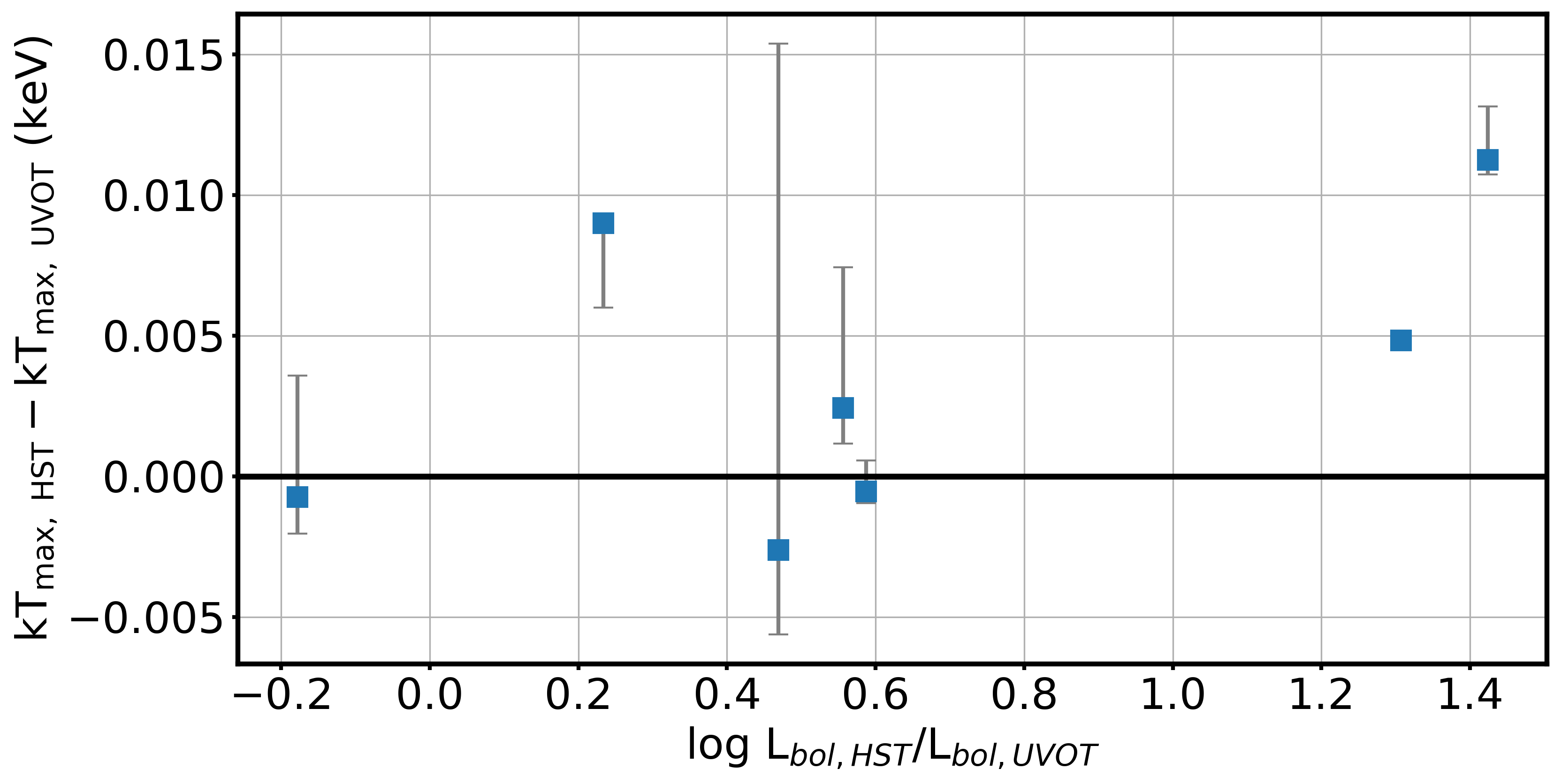}
    \caption{The difference in maximum disk temperature determined from the best fit SEDs in this work using the HST data in the UV--optical, and the maximum disk temperature from \citetalias{Gupta2024} using Swift/UVOT data as a function of the ratio between the bolometric luminosities of the two data sets.}
    \label{fig: Tmax_comp}
\end{figure}

Figure \ref{fig: Tmax_comp} shows the difference in $kT_{max}$ as a function of the change in bolometric luminosity for each source compared to that found in \citetalias{Gupta2024} using Swift/UVOT data instead of HST data. While no clear correlation is present across the all data points, the two sources with the largest differences in bolometric luminosity (log $L_{\rm bol,\;HST}/L_{\rm bol,\;UVOT} > 1.2$) show differences in the maximum disk temperature of greater than 5$\ts$eV. This large difference in bolometric luminosity seems to be driven by this change in the maximum disk temperature, which greatly increases the emission of the disk between 0.01 and 0.1$\ts$keV, where no data currently exist in either data set to constrain the fits.

The difference in the derived disk temperature is most extreme for the two sources with the lowest black hole masses, UGC3478 and Mrk1044. Due to the low black hole mass, the normalization for these sources was significantly lower than that used when fitting this source in \citetalias{Gupta2024} (which allowed the normalization to be a free parameter due to the larger number of available data points in the UV--optical). This discrepancy in the normalization helps drive the change in the disk temperature, due to the degeneracy between these two parameters. This highlights that, even with high-resolution data, there is a need for more complete SEDs spanning a wide redshift range to model AGN disks consistently and comprehensively.

\begin{figure}
    \centering
    \includegraphics[width=\linewidth]{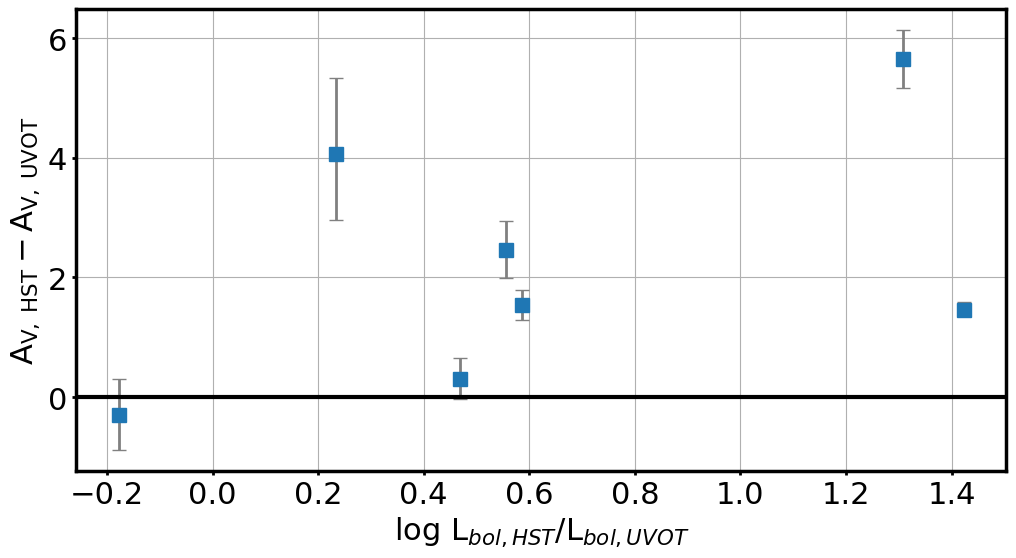}
    \caption{The difference in the extinction magnitude determined from the best fit SEDs in this work using the HST data in the UV--optical, and the extinction magnitude from \citetalias{Gupta2024} using Swift/UVOT data as a function of the ratio between the bolometric luminosities of the two data sets.}
    \label{fig: extinction_comp}
\end{figure}

\subsection{host-galaxy Dust Extinction}
One component (ZDUST) of the UV-optical SED model accounts for the intrinsic dust extinction of the source's emission due to the host-galaxy. This is quantified by the parameter E(B--V). The resulting host-galaxy dust extinction from each fit is shown in Table \ref{tab:fitting results}, and a comparison of the best-fit extinction with that found in \citetalias{Gupta2024} using the Swift/UVOT data is shown in Figure \ref{fig: extinction_comp}. For clarity, the extinction is represented in magnitude ($A_{V}$) in Figure \ref{fig: extinction_comp} instead of E(B -- V). This conversion was performed by assuming extinction properties similar to those of the SMC, as in \cite{Richards2003AJ....126.1131R,Hopkins2004AJ....128.1112H}.

Unlike the maximum disk temperature, the extinction magnitude found with the HST data was larger than that found with the Swift/UVOT data for all but one source, with a maximum difference of 5.6 magnitudes for IC2921. It is interesting to note that, for the SEDs shown in Figure \ref{fig: SEDs1}, NGC7603 shows one of the largest differences in UV emission between the HST and Swift/UVOT data (as can also be seen in Figure \ref{fig: mag comp}), but only shows a small change in the host-galaxy extinction ($\sim0.3\ts$mag). This highlights that the HST F225W wavelength does not probe the UV region where the greatest change in emission due to dust extinction occurs, and that further EUV data are critical for properly measuring the extent of UV extinction. It can also be seen that the two sources with the largest change in bolometric luminosity also show the largest increase in extinction. This highlights the importance of the contribution to the bolometric luminosity from the EUV, as well as the NIR, where significant differences in the slope of the SED can be seen, due to the addition of the constraints imposed by the inclusion of the HST F814W data point.

\begin{figure}
    \centering
    \includegraphics[width=\linewidth]{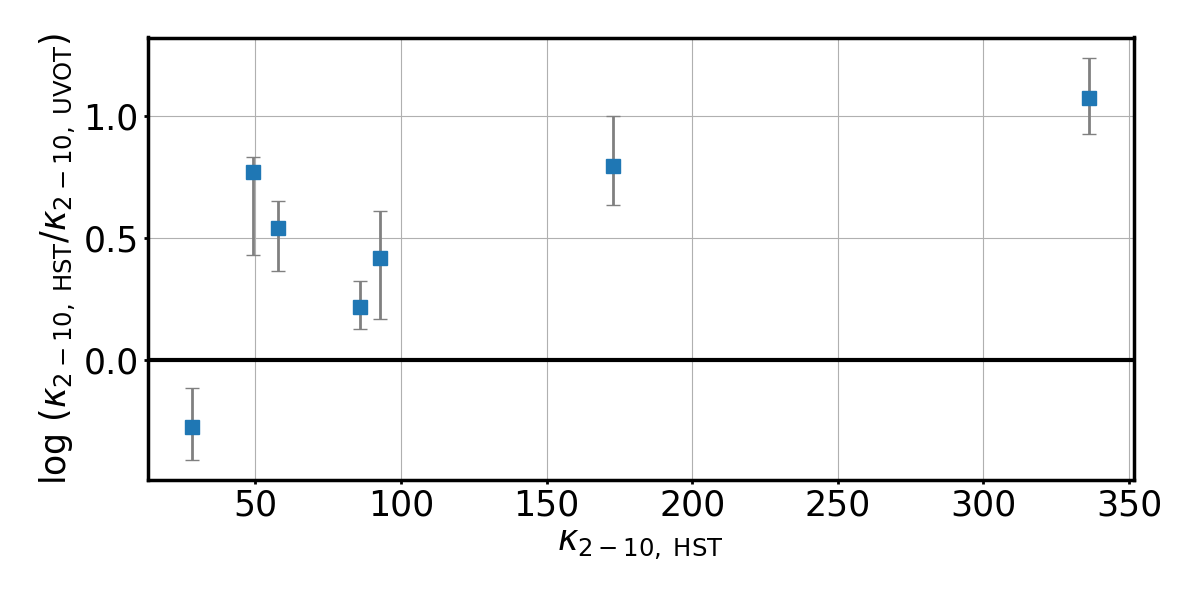}
     \caption{The difference in the bolometric corrections (HST - UVOT) from the best fit SEDs of the two data sets (HST analyzed in this work and Swift/UVOT analyzed in \citetalias{Gupta2024}) as a function of the the bolometric correction found from the HST data.}
    \label{fig: Lbol_corr_comp}
\end{figure}

\subsection{X-ray to Bolometric Correction}

Finally, we examine how the change from Swift/UVOT to higher-resolution HST data affects the bolometric correction ($\kappa_{\lambda}$). Similar to \citetalias{Gupta2024}, we calculate the bolometric correction in the X-ray by taking a ratio of the bolometric luminosity ($L_{\rm{bol}} = L_{\rm{X}} + L_{\rm{uvo}}$) and the intrinsic luminosity in the X-ray ($\kappa_{2-10\ts \rm{keV}} = L_{\rm{bol}}/L_{2-10\ts \rm{keV}}$)

Figure \ref{fig: Lbol_corr_comp} shows the change in bolometric correction from the 2--10$\ts$keV X-ray band as a function of the bolometric correction found in this work. As expected, the extent of this change matches what is seen in Figure \ref{fig: Lbol_comp}, spanning from $\sim-0.3 - 1.0$ dex. While the exact X-ray luminosity may vary slightly due to the intrinsic variability of the source and small differences in the fits (as can be seen in Figure \ref{fig:Fuvo_Fx}), the X-ray data used for the fits in this work and in \citetalias{Gupta2024} are both based on Swift/XRT observations, and the intrinsic X-ray luminosity does not differ significantly between the two fits. Therefore, the nearly entire change in bolometric luminosity and, consequently, the bolometric correction in the X-rays will be driven by the change in UV/optical luminosity between the fits using Swift/UVOT data and the HST data. 
For two sources in our sample, we find bolometric corrections of $\kappa_{\rm 2-10\;keV} > 100$. While such values are not unreasonable for sources at bolometric luminosities of log $L_{\rm bol} \approx 45.5$, as can be seen from the bolometric correction relations in works such as \cite{Duras:2020:A73}, these sources exhibit the largest deviation in the bolometric corrections and total bolometric luminosity from those found in \citetalias{Gupta2024}. This emphasizes the inherent uncertainty in measuring bolometric luminosity from limited data sets and the effect the model free parameters and model degeneracies play in the resulting best-fit SED. To better determine AGN properties, high resolution imaging data, such as that from HST used in this work, are needed to effectively disentangle AGN and host-galaxy emission. Additionally, more complete SEDs that extend into the EUV will provide vital constraints on the extent of the disk emission, helping to break the degeneracies between the model normalization and the maximum disk temperature.

\section{Conclusions} \label{sec: conclusions}
In this work, we present an analysis of the resolved UV -- optical SEDs for seven hard X-ray selected AGN from the BASS survey, utilizing high-resolution HST data. We demonstrate the power of this dataset to more effectively disentangle AGN emission from that of the host-galaxy than is possible with lower resolution data, such as that from Swift/UVOT. Each UV--optical SED was fitted with both an accretion disk model (DISKPN) and an extinction model (ZDUST) in XSPEC, mimicking the method used in previous studies that utilized Swift/UVOT data \citep{Gupta2024}.

We directly compared the HST-fitting results with those reported for \citetalias{Gupta2024} and found notable differences in the best-fit models. The two free parameters used in our models that can be directly compared to those in \citetalias{Gupta2024} are the maximum disk temperature ($kT_{\rm max}$) and the dust extinction of the AGN in the host-galaxy ($E(B-V)$). We found that the maximum disk temperature has an average difference of 0.002$\ts$keV (maximum of 0.009$\ts$keV), and the host-galaxy extinction shows an average difference of $A_{V} = 2.2$ (maximum of 5.7). These differences of the best-fit model have significant impacts on additional derived properties, such as the AGN bolometric luminosity ($L_{\rm bol} = L_{\rm X} + L_{\rm uvo}$), which has an average difference of 0.57 dex (maximum of 1.2 dex), leading to an average change in the X-ray to bolometric correction of 0.5 dex. 

While these stark differences highlight the necessity of accurately disentangling AGN emission from that of the host-galaxy when modeling an AGN's intrinsic properties, they also underscore the general difficulties in modeling AGN SEDs across different datasets and varied assumptions. As discussed in Section \ref{sec: analysis}, the UV -- optical SEDs constructed in this work have only three data points, which limits the available number of parameters that can be left free in the model. \citetalias{Gupta2024} fits the disk model to 4--6 data points per source, allowing normalization to be treated as a free parameter alongside the disk temperature and host-galaxy extinction. In this work, we fix the normalization using black hole mass measurements and make assumptions about properties such as the inclination angle and the color-to-effective temperature ratio. While we use realistic and reasonable values for these assumptions, fixing the normalization yields a key difference in the fitting techniques between \citetalias{Gupta2024} and this work. The significant differences found between the derived properties, particularly in the maximum disk temperature, are likely a combination of differences in the AGN magnitudes and the fitting techniques, as \citetalias{Gupta2024} highlighted, due to strong degeneracies between the normalization and disk temperature. It is also important to note that the model fit to the UV data in this work contains an additional component, that was not included in \citetalias{Gupta2024} in order to include photoionization absorption. While it was found that the best fit parameters do not change significantly without this model, as it primarily changes the shape of the SED in the region with no observations (0.01 -- 0.1$\ts$keV), these types of changes to the fitting process may also drive some of the differences seen between this work and \citetalias{Gupta2024}. Additional high-resolution EUV data would help break these degeneracies and better constrain disk models between 0.01 and 0.1$\ts$keV, where no observational data currently exist in either this work or \citetalias{Gupta2024}.

Going forward, the degeneracies in AGN SED modeling results should be carefully considered, and steps should be taken to ensure that the derived AGN properties are reported as accurately as possible. This includes using the best available data to isolate AGN emission and remove any extended features in the host-galaxy's bulge, as well as modeling X-ray and UV/optical emission simultaneously to reduce the limitations of fixing model parameters based on the available data. 


\begin{acknowledgments}
C.A. and M.K. acknowledge support from NASA through ADAP award 80NSSC22K1126.  

This work was performed in part at the Aspen Center for Physics, which is supported by National Science Foundation grant PHY-2210452. The authors thank Christine Done for useful discussions regarding this work.

NASA provided support for program 16241 through a grant from the Space Telescope Science Institute, which is operated by the Association of Universities for Research in Astronomy, Inc., under NASA contract NAS 5–26555.

We gratefully acknowledge funding from ANID:
CATA BASAL FB210003 (C.R., F.E.B., E.T.);
Millennium Science Initiative AIM23-0001 (F.E.B.);
and FONDECYT Regular 1241005 and 1250821 (F.E.B., E.T.).

B.T. acknowledges support from the European Research Council (ERC) under the European Union's Horizon 2020 research and innovation program (grant agreement number 950533), and by the Excellence Cluster ORIGINS which is funded by the Deutsche Forschungsgemeinschaft (DFG, German Research Foundation) under Germany's Excellence Strategy - EXC 2094 - 390783311.

KKG acknowledges financial support from the Belgian Federal Science Policy Office (BELSPO) in the framework of the PRODEX Programme of the European Space Agency.
This research is based on observations made with the NASA/ESA Hubble Space Telescope obtained from the Space Telescope Science Institute, which is operated by the Association of Universities for Research in Astronomy, Inc., under NASA contract NAS 5–26555. These observations are associated with programs 16241 and 17310.

A.T. acknowledges financial support from the Bando Ricerca Fondamentale INAF 2022 Large Grant ‘Toward a holistic view of the Titans: multiband observations of $z > 6$ QSOs powered by greedy supermassive black holes.

RS acknowledges funding from the CAS-ANID grant number CAS220016

MS acknowledges financial support from the Italian Ministry for University and Research, through the grant PNRR-M4C2-I1.1-PRIN 2022-PE9-SEAWIND: Super-Eddington Accretion: Wind, INflow and Disk-F53D23001250006-NextGenerationEU

KO acknowledges support from the Korea Astronomy and Space Science Institute under the R\&D program, supervised by the Korea AeroSpace Administration, and the National Research Foundation of Korea (NRF) grant funded by the Korea government (MSIT) (RS-2025-00553982).

DBS gratefully acknowledges support from NSF Grant 2407752.

\end{acknowledgments}


\appendix
\onecolumngrid
\renewcommand\thetable{\thesection.\arabic{table}}    
\setcounter{table}{0}

\section{Swift Data}
Table \ref{tab:swift_obs} provides the observational information for the Swift data used throughout this work, including the Swift/XRT data used in the fit with the HST data, as well as the Swift/UVOT data from \citetalias{Gupta2024} that can be seen in Figure \ref{fig: SEDs1}.

\begin{table}
\centering
\caption{Swift Observations}
\label{tab:swift_obs}

\begin{tabular}{cccc}
\hline\hline
\text{BAT ID} & \text{Counterpart Name} & \text{\textit{Swift}/XRT} & \text{\textit{Swift}/UVOT} \\
(1) & (2) & (3) & (4) \\
\hline
1189 & NGC7603 & 49538059 & 35365002 \\
335  & UGC3478 & 80373001 & 80373001 \\
34   & UGC524  & 80867001 & 80867001 \\
134  & NGC985  & 89293001 & 36530005 \\
925  & VII Zw 142 & 81213002 & 81213003 \\
130  & Mrk1044 & 35760032 & 35760002 \\
549  & IC2921  & 80057001 & 80057001 \\
\hline
\end{tabular}

\vspace{2mm}
\begin{minipage}{\linewidth}
\footnotesize
\textbf{Note.} 
(1) \textit{Swift}/BAT ID \citep{Baumgartner:2013:19}; 
(2) Counterpart galaxy name; 
(3) Observation ID of the \textit{Swift}/XRT data used in the fits shown in Fig.~\ref{fig: SEDs1}; 
(4) Observation ID of the \textit{Swift}/UVOT data used in \citetalias{Gupta2024} and shown in Fig. \ref{fig: SEDs1}.
\end{minipage}

\end{table}

\section{GALFIT Quality Checks} \label{sec: GALFIT quality check}
Here, we describe the image decomposition fitting procedure performed with GALFIT in more detail, including a discussion of the quality of the PSFs used in the fits, the fit residuals, and how the errors of the resulting point-source magnitudes are estimated. 

\subsection{Point Spread Function}
\begin{figure*}
    \centering
    \includegraphics[width=\linewidth]{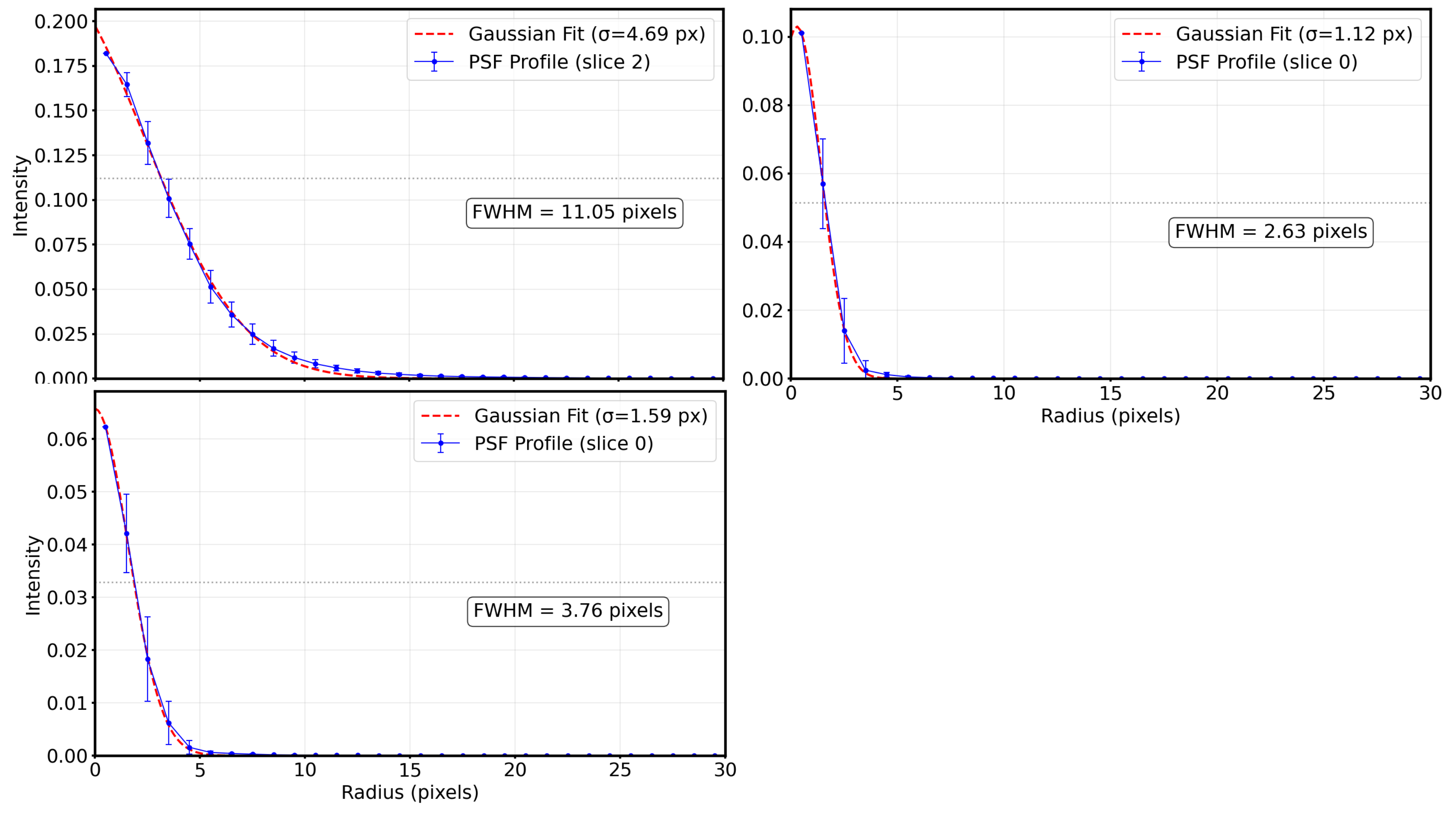}
    \caption{An example of the radial profile of the model PSF used for the F225W fits (top left), and the empirical PSFs used for the F435W (top right) and F814W (bottom left) fits.}
    \label{afig:F225W_psf}
\end{figure*}

As discussed in section \ref{subsec: GALFIT}, GALFIT requires a reliable PSF to determine accurate point source magnitudes for each fit.  We verified the quality of the PSFs used in the GALFIT analysis by comparing them to isolated, unsaturated stars in the same HST detector and filter configuration. Each empirical or synthetic PSF was fit with a two-dimensional Gaussian function to characterize its full width at half maximum (FWHM) and to confirm consistency with the expected instrumental resolution. Figure \ref{afig:F225W_psf} shows examples of these fits for the model PSF used for WFC3/UVIS F225W and the PSFs created from field stars for the ACS/WFC F435W and F814W filters. The panels display the radial intensity profile of the PSF, the best-fitting Gaussian model, and the corresponding residuals. The fitted Gaussian widths agree with the expected HST PSF FWHM values for these filters ($\sim$0.07''--0.09''), after accounting for the pixel scale and oversampling factors applied to the PSF models. The absence of significant asymmetries indicates that the PSFs are well-centered and properly sampled.

\subsection{GALFIT Residuals} \label{sec: GALFIT residuals}
\begin{figure}
    \centering
    \includegraphics[width=0.8\linewidth]{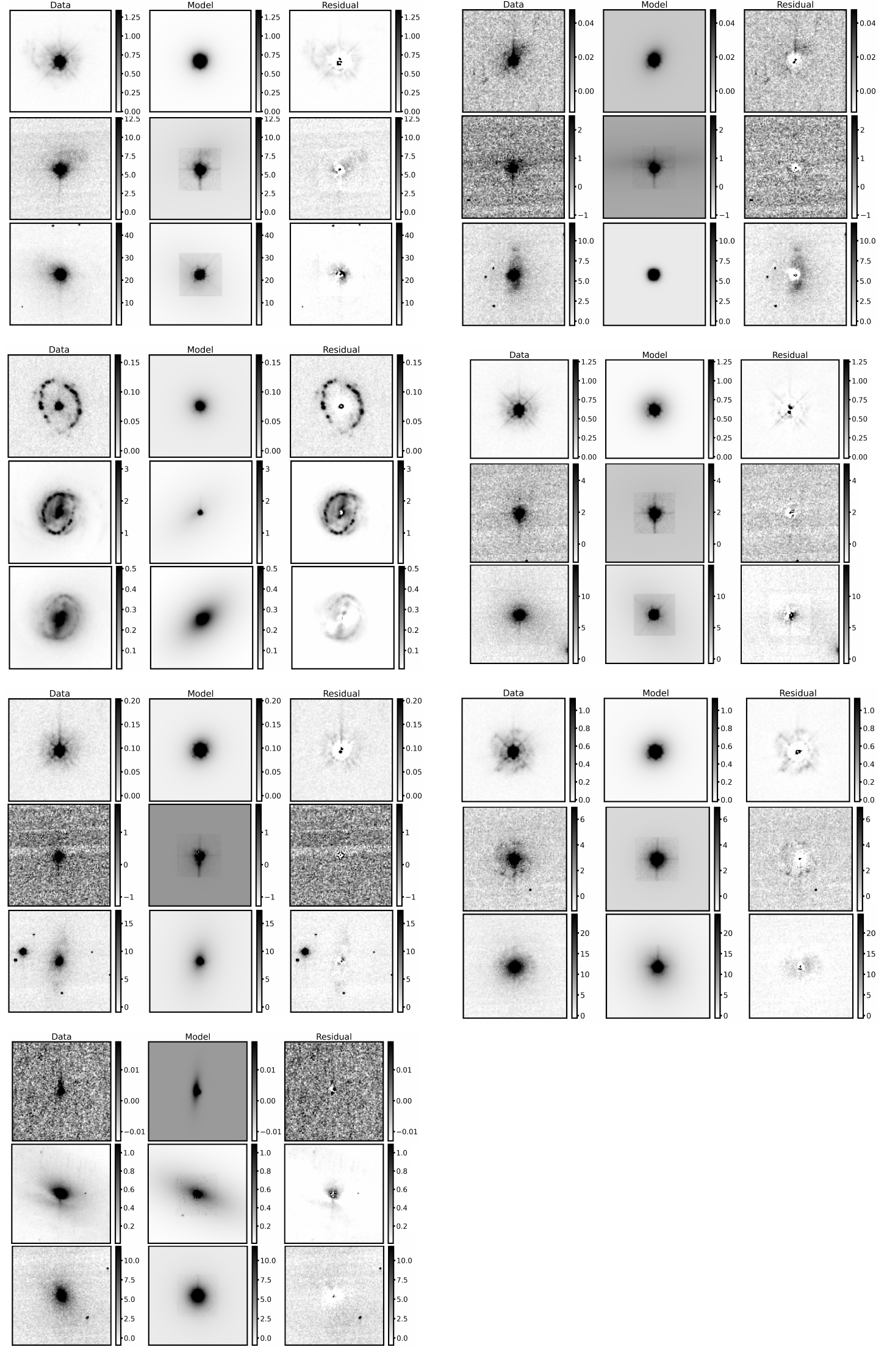}
    \caption{Results of the fits from the GALFIT analysis. Each source is grouped individually with a 3x3 grid showing the data, model, and residual (from left to right) for each filter. Each row is a different filter with F225W on the top, F435W in the middle, and F814W on the bottom. The sources are organized as: NGC7603 (upper left), UGC3478 (upper right), UGC524 (second left), Mrk1044 (second right), NGC985 (third left), VIIZw142 (third right), IC2921 (bottom left).}
    \label{fig:GALFIT_resid}
\end{figure}

Here we discuss the quality of the resulting fits from the two-dimensional surface brightness modeling performed with GALFIT. In Figure \ref{fig:GALFIT_resid}, we show the observed image, the best-fitting model, and the residual image in three filters to illustrate the quality of the fits for each source. Each panel in Figure \ref{fig:GALFIT_resid} displays, from left to right, the original HST image (‘Data’), the GALFIT model (‘Model’), and the residual (‘Data – Model’) after subtraction. All images are shown on the same spatial and intensity scales to facilitate visual comparison. The residuals highlight the fit's effectiveness and any remaining non-axisymmetric or unresolved structures. In general, the GALFIT models reproduce the central point source, while the extended host morphology, which was not the focus of this analysis, tends to exhibit larger residual structures.

\subsection{Estimating Photometric Errors} \label{sec: GALFIT errors}
Accurate photometric uncertainties for point sources fitted with GALFIT \citep{Peng:2002:266} are often underestimated, especially in the presence of non-Gaussian noise, PSF mismatch, or correlated pixel noise introduced by drizzling. To obtain more realistic magnitude uncertainties for the AGN component, we generated multiple noise realizations of the best-fit model image and refitted each realization with GALFIT. The scatter in the recovered magnitudes provides an empirical estimate of the photometric uncertainty that can be propagated into subsequent SED fitting analyses.

The procedure uses the best-fit GALFIT model image as a noise-free baseline and adds synthetic noise derived from the statistical properties of the observed HST drizzled image. This process produces a set of synthetic images that emulate the expected pixel-to-pixel noise characteristics of the data while preserving the model structure of the fitted galaxy and AGN.

\begin{figure}
    \centering
    \includegraphics[width=\linewidth]{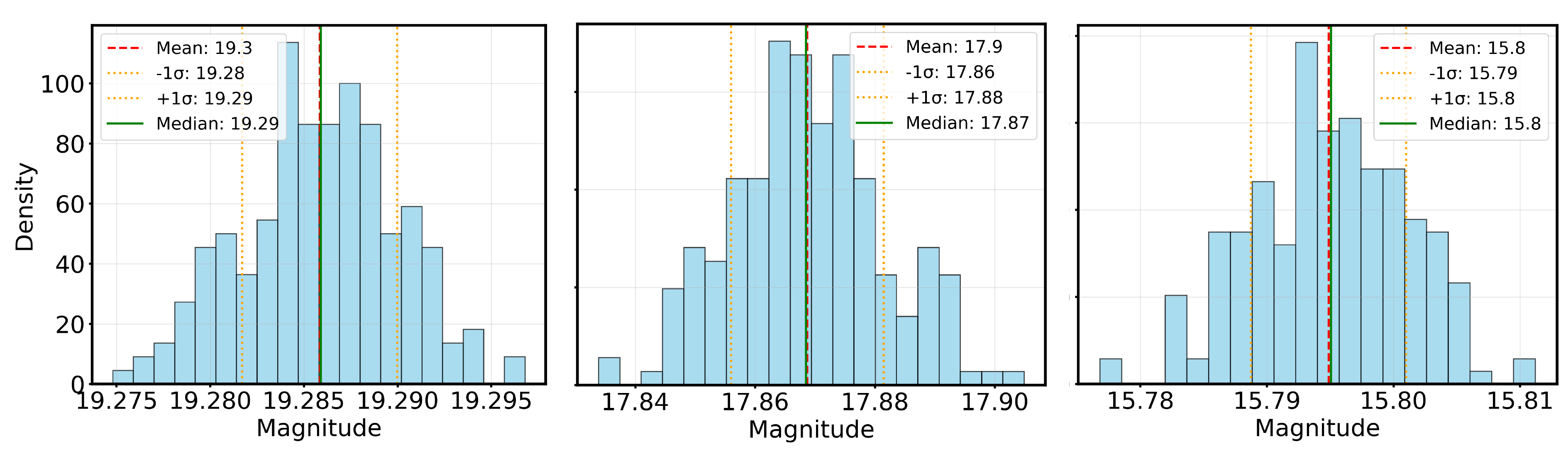}
    \caption{The results of MCMC-GALFIT analysis used to constrain the photometric errors of the best fit point source magnitudes for all three filters of UGC3478 ({\it left:} F225W {\it middle:} F435W {\it right:} F814W. This shows the point source magnitude of 500 realizations.}
    \label{fig:GALFIT_mcmc}
\end{figure}

Each realization is refitted independently with the same GALFIT configuration, and the resulting parameter distributions are used to estimate statistical uncertainties. This approach is conceptually similar to Monte Carlo Markov Chain (MCMC) sampling of the posterior distribution but is implemented through image-level noise perturbations. Figure \ref{fig:GALFIT_mcmc} shows the results of this MCMC-GALFIT analysis.

\section{UV Light Curves} \label{sec: light curves}
As an additional check for the role that variability may play in driving the differences in the measure point source magnitude between the Swift/UVOT data and HST data, we construct light curves in the Swift/UVOT UVM2 filter for all seven AGN using the available Swift monitoring data. The magnitudes are measured following the steps outlined in the Swift/UVOT data analysis guide, using region files for the source and background defined in SAO-DS9. The UVM2 images were input into the \texttt{uvotsource} command from the Swift calibration database with the two region files. This is the same methodology that was outlined in \S \ref{subsec:HST phot} to measure the magnitudes used in the comparison of the large aperture photometry.

\begin{figure}
    \centering
    \includegraphics[width=0.95\linewidth]{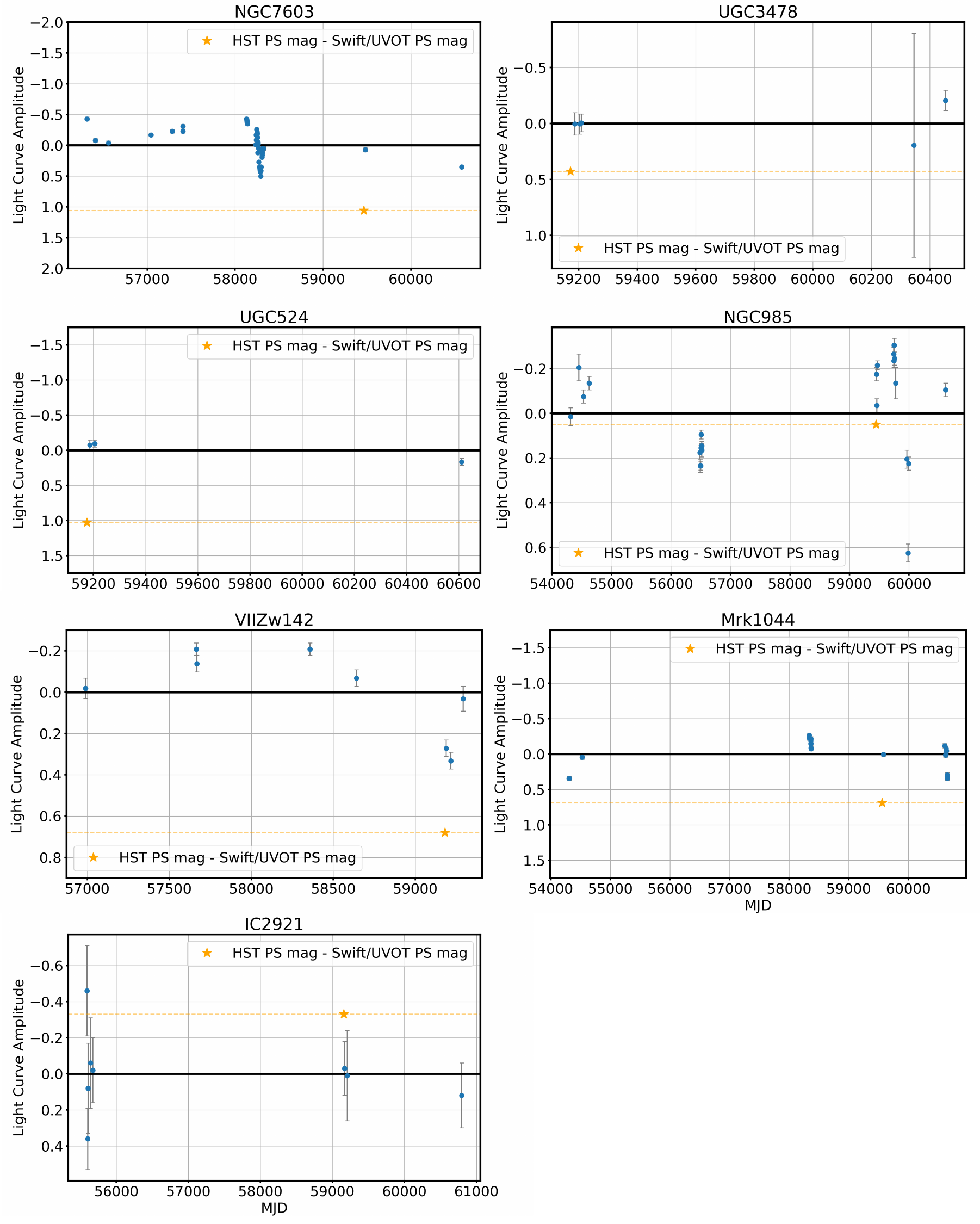}
    \caption{Swift/UVOT UVM2 light curves for the all seven BASS AGN in the sample. The light curves are constructed using large aperture photometry with a 5'' aperture. The difference in the point source magnitude measured with the HST F255W data and the Swift/UVOT UVM2 data is shown as the yellow star. For all but two sources, NGC985 and VIIZw142, the difference in the measured point source magnitude is larger than the amplitude of the light curves.}
    \label{fig:light_curves}
\end{figure}

Figure \ref{fig:light_curves} shows the amplitude of each light curve (measured magnitude -- mean magnitude of the entire light curve) along with the difference in the point source magnitude measured with the HST F225W data and that measured with the Swift/UVOT UVM2 data in \citetalias{Gupta2024}. It can be seen that, for all but one of the sources, the light curve amplitude is smaller than the difference in point source magnitudes, indicating that the systematic difference between the Swift/UVOT and HST point source magnitudes presented in \S \ref{sec: analysis} is not driven by this intrinsic variability, but is consistent with real differences in the spatial resolution and host-galaxy contamination. The source that show variability larger than the difference in AGN point source magnitude is also the source that show the smallest difference in the measured point source magnitude, NGC985. It cannot be ruled out that the difference in the measured point source magnitude or 0.05 for this source, may be driven by variability rather than due to the improvement in image resolution between the two data sets.

\facilities{HST(ACS and UVIS), Swift(XRT and UVOT)}


\bibliographystyle{aasjournalv7}


{\bibliography{new_bib}}

\end{document}